\documentclass[epj]{svjour}
\usepackage[dvips]{graphicx}
\usepackage{subeqnarray}
\usepackage{amsmath,amsfonts}
\graphicspath{{ps/}{eps/}{epsi/}{./}{../ps/}{../eps/}
		{/graal/throng/article_eta/figures/}}

\begin{document}

\title{Measurement of $\eta$ photoproduction on the proton
from threshold to 1500~MeV}

\author{
O.~Bartalini\inst{2,10},
V.~Bellini\inst{13,6},
J.P.~Bocquet\inst{1},
P.~Calvat\inst{1},
M.~Capogni\inst{2,10,4},
L.~Casano\inst{10},
M.~Castoldi\inst{8},
A.~D'Angelo\inst{2,10},
J.-P.~Didelez\inst{16},
R.~Di~Salvo\inst{10},
A.~Fantini\inst{2,10},
D.~Franco\inst{2,10},
C.~Gaulard\inst{5,14},
G.~Gervino\inst{3,11},
F.~Ghio\inst{9,12},
G.~Giardina\inst{7,17},
B.~Girolami\inst{9,12},
A.~Giusa\inst{13,7},
M.~Guidal\inst{16},
E.~Hourany\inst{16},
R.~Kunne\inst{16},
A.~Lapik\inst{15},
P.~Levi~Sandri\inst{5},
A.~Lleres\inst{1},
F.~Mammoliti\inst{13,7},
G.~Mandaglio\inst{7,17},
D.~Moricciani\inst{10},
A.N.~Mushkarenkov\inst{15},
V.~Nedorezov\inst{15},
L.~Nicoletti\inst{2,10,1},
C.~Perrin\inst{1},
C.~Randieri\inst{13,6},
D.~Rebreyend\inst{1},
F.~Renard\inst{1},
N.~Rudnev\inst{15},
T. Russew\inst{1},
G. Russo\inst{13,7},
C.~Schaerf\inst{2,10},
M.-L.~Sperduto\inst{13,7},
M.-C.~Sutera\inst{7},
A.~Turinge\inst{15},
V.~Vegna\inst{2,10}
(The GRAAL Collaboration)
}
\offprints{lleres@lpsc.in2p3.fr}   
\institute{LPSC, Universit\'e Joseph Fourier Grenoble 1, CNRS/IN2P3, Institut National Polytechnique de Grenoble, 53 avenue des Martyrs, 38026 Grenoble, France
\and 
Dipartimento di Fisica, Universit\`a di Roma "Tor Vergata", via della Ricerca Scientifica 1, I-00133 Roma, Italy
\and
Dipartimento di Fisica Sperimentale, Universit\`a di Torino, via P. Giuria, I-00125 Torino, Italy
\and
Present affiliation: ENEA - C.R. Casaccia, via Anguillarese 301, I-00060 Roma, Italy
\and
INFN - Laboratori Nazionali di Frascati, via E. Fermi 40, I-00044 Frascati, Italy
\and
INFN - Laboratori Nazionali del Sud, via Santa Sofia 44, I-95123 Catania, Italy
\and
INFN - Sezione di Catania, via Santa Sofia 64, I-95123 Catania, Italy
\and
INFN - Sezione di Genova, via Dodecanneso 33, I-16146 Genova, Italy
\and
INFN - Sezione di Roma, piazzale Aldo Moro 2, I-00185 Roma, Italy
\and
INFN - Sezione di Roma Tor Vergata, via della Ricerca Scientifica 1, I-00133 Roma, Italy
\and
INFN - Sezione di Torino, I-10125 Torino, Italy
\and
Istituto Superiore di Sanit\`a, viale Regina Elena 299, I-00161 Roma, Italy
\and
Dipartimento di Fisica ed Astronomia, Universit\`a di Catania, via Santa Sofia 64, I-95123 Catania, Italy
\and 
Present affiliation: CSNSM, Universit\'e Paris-Sud 11, CNRS/IN2P3, 91405 Orsay, France
\and
Institute for Nuclear Research, 117312 Moscow, Russia
\and
IPNO, Universit\'e Paris-Sud 11, CNRS/IN2P3, 15 rue Georges Cl\'emenceau, 91406 Orsay, France
\and
Dipartimento di Fisica, Universit\`a di Messina, salita Sperone, I-98166 Messina, Italy}

\date{Received: date / Revised version: date}
\abstract{Beam asymmetry and differential cross section for the reaction
$\gamma$p$\rightarrow\eta$p were measured from production threshold to 1500~MeV 
photon laboratory energy. The two dominant neutral decay modes of the $\eta$ meson,
$\eta\rightarrow 2\gamma$ and $\eta\rightarrow 3\pi^0$, were analyzed. 
The full set of measurements is in good agreement 
with previously published results. Our data were compared with three models.
They all fit satisfactorily the results but their respective resonance contributions are
quite different. The possible photoexcitation of a narrow state N(1670) was 
investigated and no evidence was found.} 
\maketitle
\section{Introduction}
\label{intro}

Eta photoproduction on the proton in the resonance region has been abundantly studied over 
the last years \cite{kru95}-\cite{els07} and the initial 
expectation of a simple reaction mechanism has faded away. For the time being, 
apart from the well-established contributions of two resonances, the dominant $S_{11}$(1535) 
and the $D_{13}$(1520) whose excitation was clearly revealed by our beam asymmetry measurement 
close to threshold \cite{aja98}, the contribution of states in the third resonance region 
remains largely model-dependent \cite{li98}-\cite{sar05}. Some models even need to 
incorporate new resonances \cite{sag01},\cite{ani05}. 

Presently, the $\eta$ photoproduction database contains mostly cross section results and 
only a few single polarization observable data. In addition to our beam asymmetry measurement, 
the target asymmetry was measured at the Bonn synchrotron up to 1100~MeV \cite{boc98} and
some preliminary beam asymmetries have been recently obtained by the CB-ELSA/TAPS collaboration
up to 1350~MeV \cite{els07}. Polarization observables, being sensitive to 
interference terms between different multipoles, 
bring valuable constraints on partial wave analyses and therefore it is 
desirable to extend these measurements in the third resonance region.

In the present work, we report on precise measurements of the beam asymmetry $\Sigma$
and of the differential cross section for the reaction $\gamma$p$\rightarrow
\eta$p from production threshold (E$_{\gamma}$=707 MeV) to 1500~MeV (W=1485-1900~MeV). 
The extracted total cross section is also presented. This work complements
and improves our previously published results for energies up to 1100~MeV 
\cite{aja98,ren02}.
 
\section{Experimental set-up}
\label{setup}

The experiment was carried-out with the GRAAL facility (see \cite{bar05} 
for a detailed description and references therein), installed at the European Synchrotron 
Radiation Facility (ESRF) in Grenoble (France). The tagged and polarized $\gamma$-ray 
beam is produced by Compton scattering of laser photons off the 6.03~GeV 
electrons circulating in the storage ring.

In the present experiment, we used alternately the green line
at 514~nm and a set of UV lines around 351~nm produced by an Ar 
laser, giving 1.1 and 1.5~GeV $\gamma$-ray maximum energies, respectively.

The photon energy is provided by an internal tagging system consisting
of silicon microstrips (128 strips with a pitch of 300~$\mu m$) for measurements
of the scattered electron position 
and a set of plastic scintillators for Time-of-Flight (ToF) measurements. 
The measured energy resolution of 16~MeV is dominated
by the energy dispersion of the electron beam (14 MeV - all resolutions are given as FWHM). 
The energy calibration is extracted run by run from the fit of the Compton edge position with a
precision of $\sim$10$\mu m$ \footnote{This high accuracy has allowed us to improve by
three orders of magnitude the limit for the light speed anisotropy 
($\Delta c/c \leq 3 \times 10^{-12}$) \cite{gur05}}, equivalent to
$\Delta E_\gamma/E_\gamma \simeq 2 \times 10^{-4}$ (0.3~MeV at 1.5~GeV).

The energy dependence of the $\gamma$-ray beam polarization was determined 
using the Klein-Nishina formula and taking into account the laser and electron beam 
emittances. The $\gamma$-ray beam polarization is close to 100\% at the maximum 
energy and decreases smoothly with energy down to a minimum of $\approx$30\% (UV) 
or $\approx$60\% (green) at the $\eta$ production threshold. Based on detailed studies \cite{bar05}, 
it was found that the only significant source of error for the $\gamma$-ray 
polarization comes from the laser beam polarization ($\delta P_\gamma / P_\gamma$=2\%).

A thin monitor is used to measure the beam flux (typically 10$^6$ $\gamma$/s). The monitor 
efficiency (2.68$\pm$0.03\%) was estimated by comparison with the response at low rate of
a lead/scintillating fiber calorimeter. 

The target cell consists of an aluminum hollow cylinder of 4~cm in diameter 
closed by thin mylar windows (100~$\mu$m) at both ends. The target (6~cm long for the
present experiment) was filled by liquid hydrogen at 18~K ($\rho \approx 7 \ 10^{-2}$~g/cm$^3$).

\begin{figure}
\begin{center}
\includegraphics[width=0.9\linewidth]{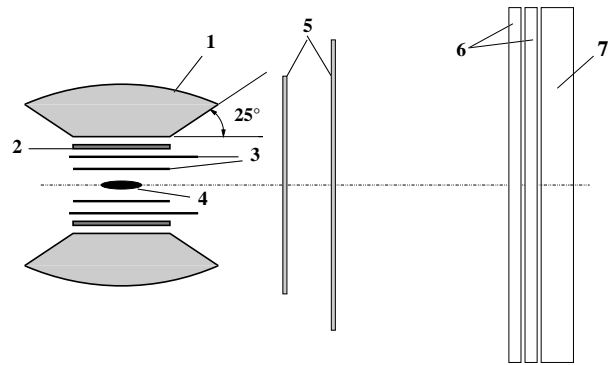} 
\end{center}
\caption{Schematic view of the LA$\gamma$RANGE detector: BGO calorimeter (1), Plastic scintillator barrel (2),
Cylindrical MWPC's (3), Target (4), Plane MWPC's (5), Double plastic scintillator hodoscope (6),
Lead-scintillator shower detector (7) (the drawing is not to scale).}
\label{sch}
\end{figure}

The 4$\pi$ LA$\gamma$RANGE detector of the GRAAL set-up allows to detect both neutral 
and charged particles (fig. \ref {sch}).

The $\gamma$-rays coming from the $\eta$ neutral decay channels ($\eta \rightarrow 2\gamma$ and 
$\eta \rightarrow 3\pi^{0} \rightarrow 6\gamma$ - branching ratios of 39.2 and 32.2\%, 
respectively) are detected in a BGO calorimeter made of 480 ($15 \theta \times 32 \varphi$) 
crystals, each of 21 radiation lengths. They are identified as clusters of adjacent crystals 
(3 on average for an energy threshold of 10 MeV per crystal) with no associated hit in the 
barrel. The measured photon energy resolution is 3\% on average.
For a thin target (3~cm), the angular resolution is 
6$^0$ and 7$^0$ for polar and azimuthal angles, respectively .

At forward angles, the $\gamma$-rays can be detected in a lead-scintillator 
sandwich ToF wall, consisting of 16 vertical modules. This detector provides a good angular 
resolution but no energy measurement and, for the present reaction, extends only marginally
the covered angular range. For the sake of simplicity, it was not used in the present
analysis.

The recoil proton track is measured by a set of MultiWire Proportional Chambers 
(MWPC) (see \cite{lle07} for more details). Two cylindrical 
chambers with striped cathodes are used to cover the central region and give a 
reconstruction efficiency $\geq$90\% with a resolution of 3.5$^0$ in $\theta$ 
and 4.5$^0$ in $\varphi$. The forward angle tracks are measured 
by two planar chambers (efficiency $\geq$ 99\%), each composed of two wire planes; the 
average polar and azimuthal resolutions are 1.5$^0$ and 2$^0$, respectively.

Charged particle identification in the central region is obtained by dE/dx technique 
thanks to a plastic scintillator barrel (32 bars, 5~mm thick, 43~cm long) with an 
energy resolution $\approx$20\%. For the charged particles emitted in the forward
direction, a Time-of-Flight measurement is provided by a double plastic
scintillator hodoscope (300$\times$300$\times$3~cm$^3$) placed at a distance of 
3~m from the target and having a resolution of $\approx$600~ps. This detector 
provides also a measure of the energy loss dE/dx. Energy calibrations were extracted
from the analysis of the $\pi^0 p$ photoproduction reaction while the ToF calibration 
of the forward wall was obtained from fast electrons produced in the target.

For the cross section measurements, due to large uncertainties on the cylindrical chambers
efficiency, the proton direction was deduced from the association between the
scintillator barrel and the BGO calorimeter at the cost of a worse resolution
($\sim 10^0$ in $\theta$ and $\varphi$).

\section{Data analysis}
\label{analysis}

\subsection{Channel selection}

For the present results, the same selection method used in our previous publications on
$\pi^0$ and $\eta$ photoproduction \cite{aja98,ren02,bar05,ren99} was applied. Only
the main points will be recalled in the following.

The analysis method is based on two-body kinematics. Thanks to the complete detection of 
all final-state products, the kinematics of the reaction is overdetermined and a clean 
event selection can be achieved without the need for background subtraction.

Only events with two or six neutral clusters in the BGO calorimeter
and a single charged-particle track were selected. Channel selection was achieved by 
applying cuts on the following quantities:

\begin{itemize}

\item[.] $M_{2\gamma}$ or $M_{6\gamma}$
\vspace{0.3cm}
\item[.] $R_{\eta}$ = $E_{\eta}/E_{\eta}^*$
\vspace{0.3cm}
\item[.] $\Delta\theta_p$ = $\theta_p^*$ - $\theta_p$
\vspace{0.3cm}  
\item[.] $\Delta\varphi_p$ = $\varphi_p^*$ - $\varphi_p$ 
\vspace{0.3cm}
\item[.] $\Delta t_p$ = $ToF_p^*$ - $ToF_p$ (only at forward angles)

\end{itemize}

\noindent where the "$^*$" indicates variables calculated from the two-body kinematics as 
opposed to measured ones. $M_{2\gamma}$ and $M_{6\gamma}$ are the invariant masses
of the detected photons. 

A Monte Carlo simulation of the apparatus based on the GEANT3 package,
coupled with a complete event generator including all known photoproduction reactions \cite{maz94},
was used to optimize selection cuts, calculate detection efficiencies and estimate
background contamination. To optimize event selection, experimental and simulated distributions 
were compared for all kinematical variables. A strong background rejection
together with a good efficiency could be achieved with cuts at $\pm$3$\sigma$. 

\begin{figure}
\begin{center}
\includegraphics[width=0.99\linewidth]{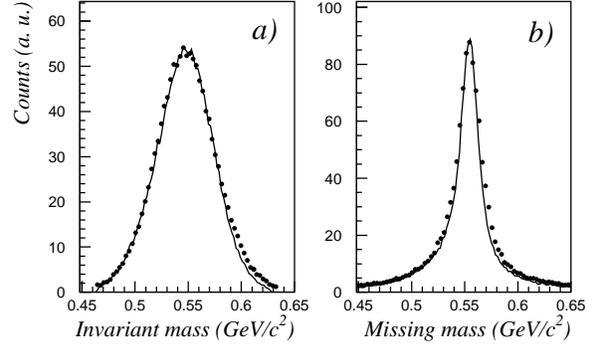} 
\end{center}
\caption{(a) Invariant mass spectrum for $\eta \rightarrow 2\gamma$ ; (b) Missing mass spectrum
calculated from the proton momentum. Data (closed circles) and simulation of the $\eta$p channel (solid line) are
compared with all kinematical cuts applied.}
\label{masse}
\end{figure}

Two examples of experimental distributions are given in fig. \ref{masse} with the invariant 
mass of the $\eta$ decaying in two $\gamma$-rays and the missing mass calculated from the 
recoil proton momentum; they are compared with what is expected from the simulation of the
$\eta$p channel. For both quantities, with all kinematical cuts applied, an overall satisfactory 
agreement is achieved, despite some slight discrepancies. These are attributed to small misalignments 
of the apparatus (beam, target, wire chambers, ...) not fully taken into account in the simulation. 
Indeed, similar discrepancies remain when varying cuts from $\pm$3$\sigma$ to $\pm$2$\sigma$,
excluding therefore a significant background contribution.

The level of hadronic background, estimated from the simulation of all possible
final states, does not exceed 1\% at 1 GeV
and increases up to 5\% at 1.5 GeV. 
Such a limited contamination was confirmed by the good agreement of the asymmetries and 
differential cross sections extracted independently for the two neutral $\eta$ decay modes
(see sect. \ref{decays}).

\subsection{Measurement of $\Sigma$}

The beam asymmetry $\Sigma$ was determined from the standard expression:

\noindent
\begin{eqnarray} 
\frac{\tilde{N}_V(\varphi)-\tilde{N}_H(\varphi)}{\tilde{N}_V(\varphi)+\tilde{N}_H(\varphi)}
= P_{\gamma}\Sigma \cos (2\varphi)
\label{rap}
\end{eqnarray} 

\noindent
where $\tilde{N}_V$ and $\tilde{N}_H$ are the azimuthal yields normalized 
by the integrated flux for the vertical and horizontal polarization states,
respectively. $P_{\gamma}$ is the degree of linear polarization of the
beam and $\varphi$ the azimuthal angle of the reaction plane. For a given
bin in energy $E_\gamma$ and $\theta_{cm}$, with $\theta_{cm}$ the $\eta$ center-of-mass angle, 
the beam asymmetry $\Sigma$ was extracted from the fit of the normalized ratio (eq. \ref{rap}) 
by the function $P_{\gamma}\Sigma\cos(2\varphi)$, using the known energy dependence of $P_{\gamma}$.
The measured asymmetries were corrected for the finite $\varphi$ binning
($\Sigma_{true}=\Sigma_{meas}(1+R_{\varphi})$ with $R_{\varphi}$=0.026 for 16 bins). 

Two sources of systematic errors were considered: i) the uncertainty on the beam polarization 
($\delta \Sigma/\Sigma$=$\delta P_\gamma / P_\gamma$
=2\%) and ii) the background contamination. 
For the second one, two main contributions were identified: other photoproduction (hadronic) 
reactions and target wall events.
The uncertainty due to hadronic contamination was estimated from the 
variation of the extracted asymmetries when opening cuts from $\pm$3$\sigma$ to $\pm$4$\sigma$. 
The resulting errors range from $\delta \Sigma$=0.003 to 0.035.
The rate of target wall events was measured via empty target runs and
found to be less than 1\%. 
The corresponding error was neglected.
All systematic and statistical errors were summed 
quadratically. The global statistical/systematic ratio was found to be of the order of 1.5.

\subsection{Measurement of $d\sigma / d\Omega$}
\label{dsdo}

The differential cross section for a given bin in $E_\gamma$
and $\cos\theta_{cm}$ was calculated using the following expression:

\noindent
\begin{eqnarray} 
\frac{d\sigma}{d\Omega}(\cos\theta_{cm},E_\gamma)&=&\frac{N(\cos\theta_{cm},E_\gamma)}{b_{\eta} \epsilon(\cos\theta_{cm},E_\gamma) F(E_\gamma) \rho l \Delta \Omega}
\end{eqnarray}

\noindent where N is the number of selected events, $b_{\eta}$ the branching ratio, $\epsilon$ the detection efficiency, 
$F$ the integrated beam flux, $\rho$ the hydrogen density, $l$ the target length and 
$\Delta \Omega$ the solid angle (in the present case $\Delta\Omega=0.2\pi$, corresponding to
20 bins in $\cos\theta_{cm}$).

The detection efficiency $\epsilon$ was derived from the simulation. The global 
efficiency, including acceptance, detection, identification and selection,
is of the order of 33\% for $\eta \rightarrow 2\gamma$ and 6\% for 
$\eta \rightarrow 3\pi^0$.

Since cross section data were obtained by summation of a large number of successive periods,
the corresponding experimental configurations were implemented in the simulation
to calculate the efficiency. In particular, special care was taken of the longitudinal 
position of the target, measured with the cylindrical MWPCs \cite{bar05}, a crucial parameter 
for the control of the acceptance.

Two types of systematic errors were taken into account: global and bin-dependent ones. 
The former type includes the uncertainties on beam flux monitor efficiency, hydrogen density 
and target length. The quadratic sum of these different contributions gives a global normalization error of 2.3\%.
The latter type takes into account uncertainties on longitudinal target position, efficiency
and hadronic background contamination. The errors corresponding to the target position
strongly depend on the bin; they are in general low ($\leq$2\%) and can reach up to 10\% for a few points.
The error due to hadronic contamination, together with the error on efficiency, 
were estimated from the variation of the extracted cross sections when opening cuts from $\pm$3$\sigma$ 
to $\pm$4$\sigma$. The resulting uncertainties 
(angular averaged) steadily increase from around 4\% at 1 GeV up to 13\% at 1.5 GeV.
Only the bin-dependent errors were summed quadratically with the statistical errors.
The global statistical/systematic ratio was found to be of the order of 1.1.

\section{Results and discussions}
\label{results}

The complete set of asymmetry and cross section data ($\sim$1~million selected $\eta$p events) 
covers large photon energy (from 700 to 1500 MeV) and $\eta$ angular 
($\theta_{cm}$=30-160$^0$) ranges. The results are displayed in figs. \ref{asym_comp} to \ref{sect2}. 
Numerical values are listed in tables \ref{table_ass1} to \ref{table_sect2}. 
For cross sections, the global normalization uncertainty of 2.3\% 
has not been included in the tabulated values nor in the plotted errors. 
The total cross section was also extracted and is plotted in fig.~\ref{sect_tot}.
Up to 850 MeV, the presented cross sections are those previously published 
which were obtained with a 3~cm long target better suited for the detection of low energy protons.

\subsection{Comparison to previous GRAAL results}

GRAAL results were already published for $\gamma$-ray energies up to 1100 MeV
($\Sigma$ \cite{aja98} and $d\sigma / d\Omega$ \cite{ren02}). The newly analyzed sample 
not only extends the energy range up to 1500 MeV but also increases tenfold the statistics.

\begin{figure}
\begin{center}
\includegraphics[width=1.0\linewidth]{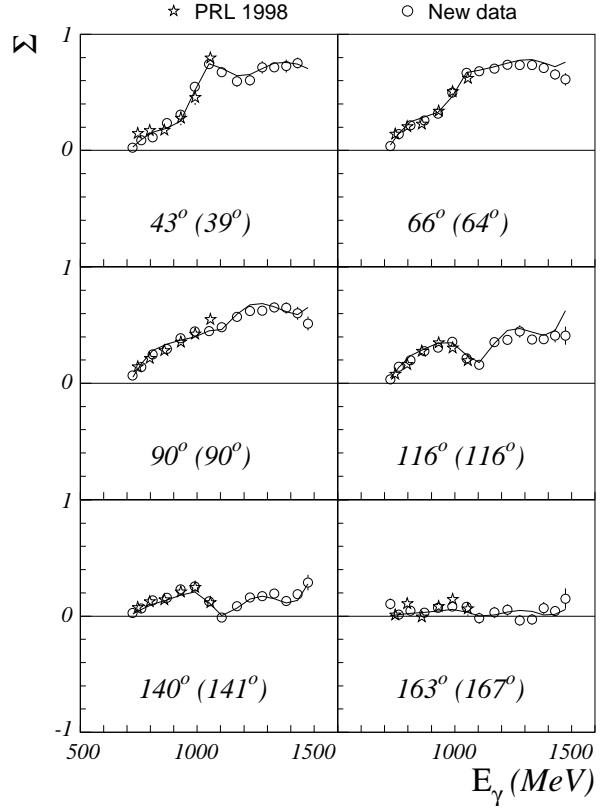} 
\end{center}
\caption{Beam asymmetry as a function of the $\gamma$-ray energy 
for different $\eta$ center-of-mass angles. Results published in 1998 up to 1100 MeV 
(stars - angle in parentheses) are compared to the new results (circles).
The curve represents the result of the BCC model (see sect. \ref{disc}).}
\label{asym_comp}
\end{figure}

\begin{figure}
\begin{center}
\includegraphics[width=1.05\linewidth]{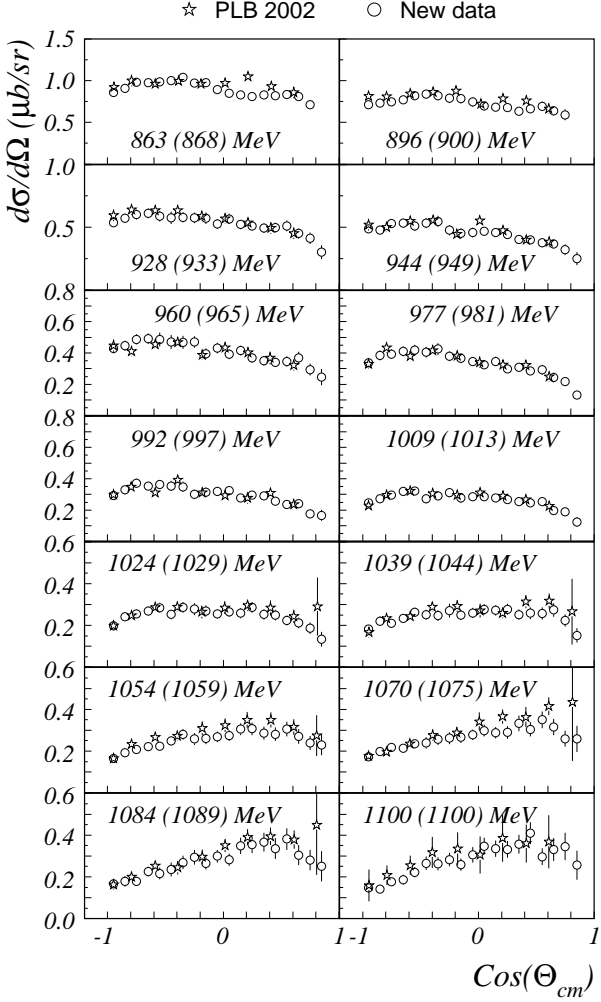} 
\end{center}
\caption{Differential cross section for photon energies ranging from 850 to 1100~MeV. Results published
in 2002 (stars - energy in parentheses) are compared to the new results (circles).}
\label{sectcg}
\end{figure}

The new results are compared with the published data in figs. \ref{asym_comp} 
($\Sigma$) and \ref{sectcg} ($d\sigma/d\Omega$ for E$_\gamma\geq$850~MeV). The agreement between the two sets
is good at all energies and angles. For the beam asymmetries, it should be remembered
that the beam polarization depends upon the energy and the used laser line. Hence, 
at a given energy, the beam polarization differs for the  UV and green laser lines;
for instance, at 1 GeV, $P_\gamma$ $\simeq$100\% for green and $\simeq$70\% for UV.
The excellent agreement confirms the good control of the beam polarization. For the 
cross sections, the comparison is even more stringent.
Indeed, these absolute measurements necessitate an accurate knowledge of the 
flux and efficiency; on top of that, the summation over numerous periods requires a precise 
monitoring of the detector response. Again, the observed good agreement between the two 
data sets demonstrates the reliability of the present analysis.

These new results improve our previous measurements particularly at forward angles and allow
to better describe the behaviour of the cross section in this angular domain.

\subsection{Comparison between the two $\eta$ neutral decays}
\label{decays}

\begin{figure}
\begin{center}
\includegraphics[width=1.05\linewidth]{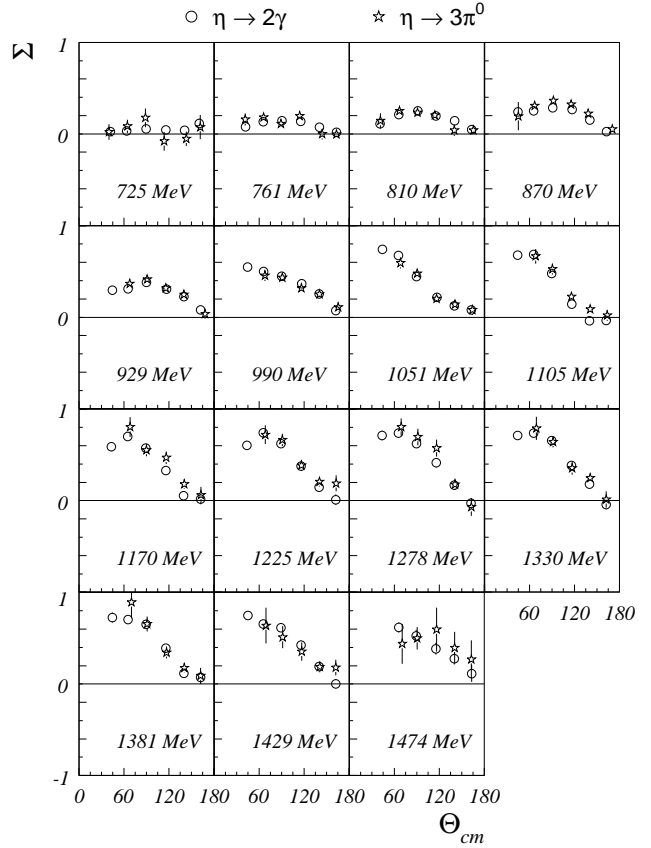} 
\end{center}
\caption{Angular distributions of the beam asymmetry. Comparison between the two neutral decay modes:
$\eta \rightarrow 2\gamma$ (circles) and $\eta \rightarrow 3\pi^{0}$ (stars).}
\label{ass26}
\end{figure}

\begin{figure}
\begin{center}
\includegraphics[width=1.05\linewidth]{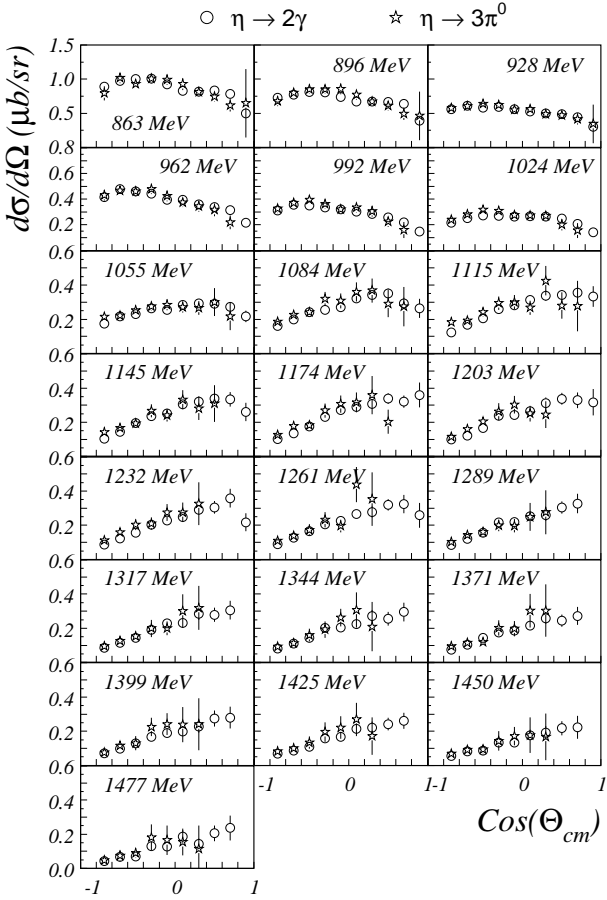} 
\end{center}
\caption{Differential cross section for energies ranging from 850 to 1500 MeV. Comparison between 
the two neutral decay modes: $\eta \rightarrow 2\gamma$ (circles) and $\eta \rightarrow 3\pi^{0}$ (stars).}
\label{sect26}
\end{figure}

The beam asymmetries and differential cross sections were calculated independently for the
two neutral decay modes, $\eta \rightarrow 2\gamma$ and $\eta \rightarrow 3\pi^{0}$
and the comparison is displayed in figs. \ref{ass26} ($\Sigma$) and \ref{sect26} ($d\sigma / d\Omega$). 
To reduce statistical errors associated with the 3$\pi^{0}$ decay, a broader angular binning was
used for the comparison.

Because the detection of the six decay photons is requested in the 3$\pi^{0}$ analysis,
the global efficiency is strongly reduced as compared to the 2$\gamma$ (see sect. \ref{dsdo})  
(the angular range is also limited to $\theta_{cm}\geq 60^0$). On the other hand, this criterion  
largely excludes the two main hadronic backgrounds ($\pi^{0}$ and 2$\pi^{0}$).
The excellent agreement observed for both quantities between the 2$\gamma$  and 3$\pi^{0}$ results
confirms the low level of background in the 2$\gamma$ channel.
In addition, the very good agreement of the cross section data demonstrates the reliability of the 
Monte Carlo simulation, especially for the cluster reconstruction in the BGO calorimeter.

\subsection{Comparison to CLAS, CB-ELSA and LNS-GeV-$\gamma$ results}

Differential cross section results have been recently published by the CLAS \cite{dug02}, CB-ELSA \cite{cre05} 
and LNS-GeV-$\gamma$ \cite{nak06} collaborations. Whereas the GRAAL data (as well as LNS-GeV-$\gamma$ results)
are absolute measurements, for the CLAS and CB-ELSA results, the 
normalization was obtained by using the SAID partial-wave analysis. 

\begin{figure}
\begin{center}
\includegraphics[width=0.9\linewidth]{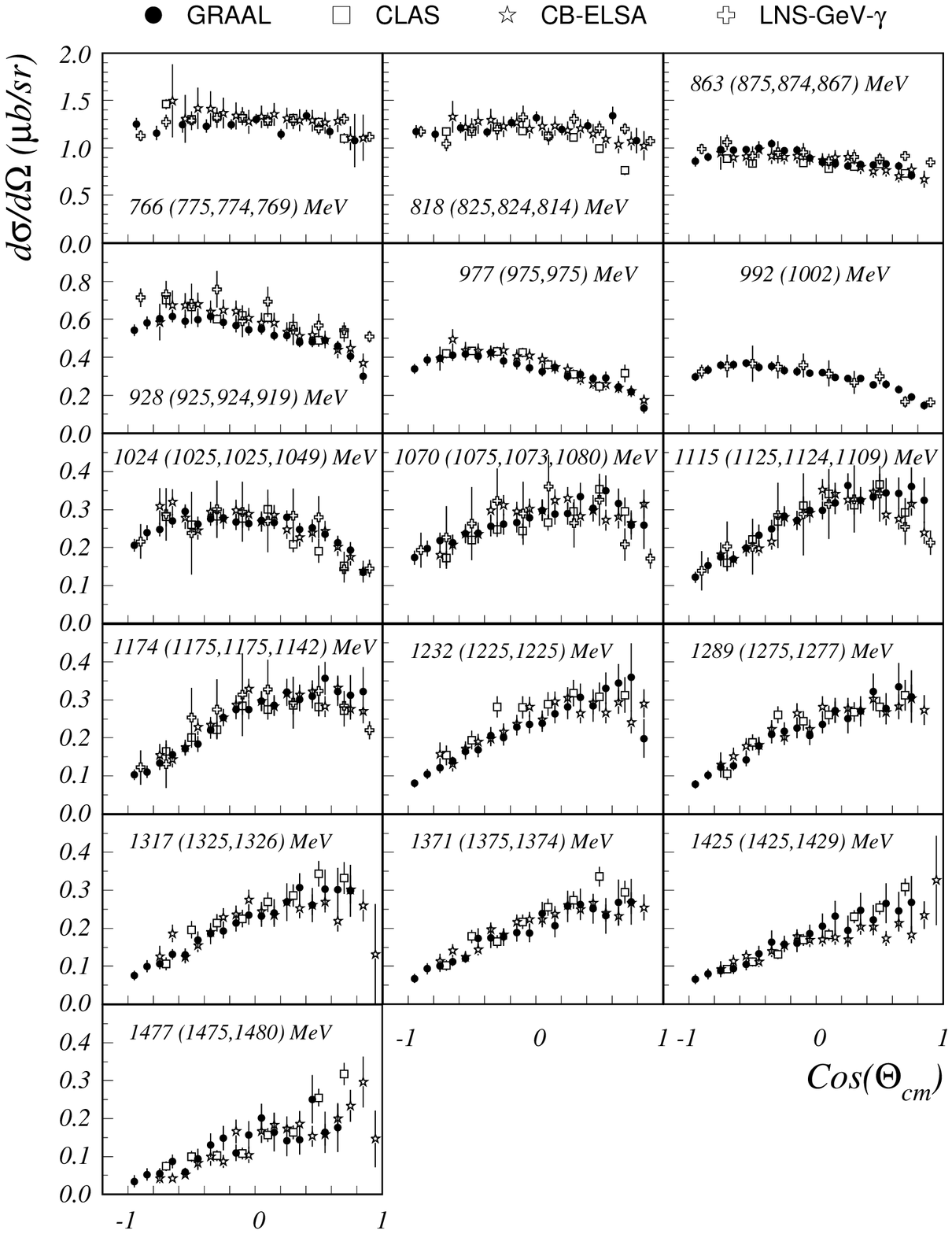} 
\end{center}
\caption{Comparison between GRAAL (closed circles), CLAS (open squares), CB-ELSA (open stars) and LNS-GeV-$\gamma$ (open crosses)
differential cross-section data for the closest energy bins of the four experiments, from threshold to 1500 MeV
(CLAS, CB-ELSA and LNS energy values are in parentheses).}
\label{sect_comp}
\end{figure}

The comparison between GRAAL, CLAS, CB-ELSA and LNS-GeV-$\gamma$ data is shown in 
fig. \ref{sect_comp} for the closest energy bins. The overall agreement is good over the whole 
energy and angular ranges. It is worth noting that our angular range is complementary to the CLAS
and CB-ELSA ones, extending the measurement to more backward angles. 

\begin{figure}
\begin{center}
\includegraphics[width=0.8\linewidth]{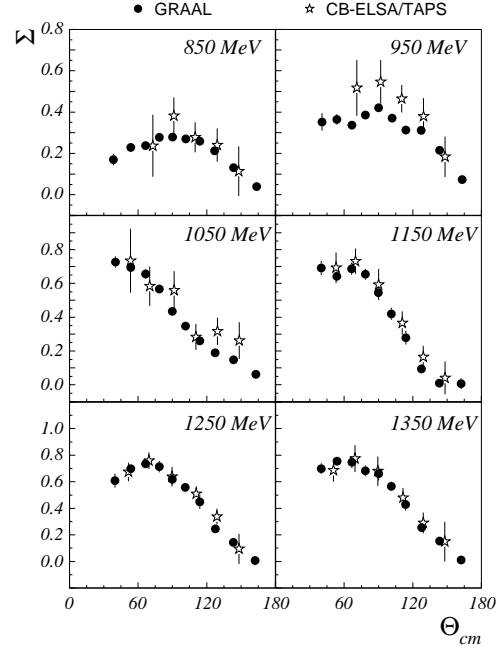} 
\end{center}
\caption{Comparison between GRAAL (closed circles) and CB-ELSA/TAPS (open stars) beam asymmetry data
for the Bonn energy bins ($\pm$50 MeV).}
\label{asym_bonn}
\end{figure}

Preliminary beam asymmetries have also been recently presented by the CB-ELSA/TAPS collaboration in the energy range
E$_\gamma$=800-1400 MeV \cite{els07}. The measurements were performed using a linearly polarized tagged 
photon beam produced by coherent bremsstrahlung off a diamond. A nice agreement is found with our data, 
except at 950~MeV where sizeable discrepancies are observed (fig. \ref{asym_bonn}).

\subsection{Total cross section}
\label{tot_sect}

\begin{figure}
\begin{center}
\includegraphics[width=1.05\linewidth]{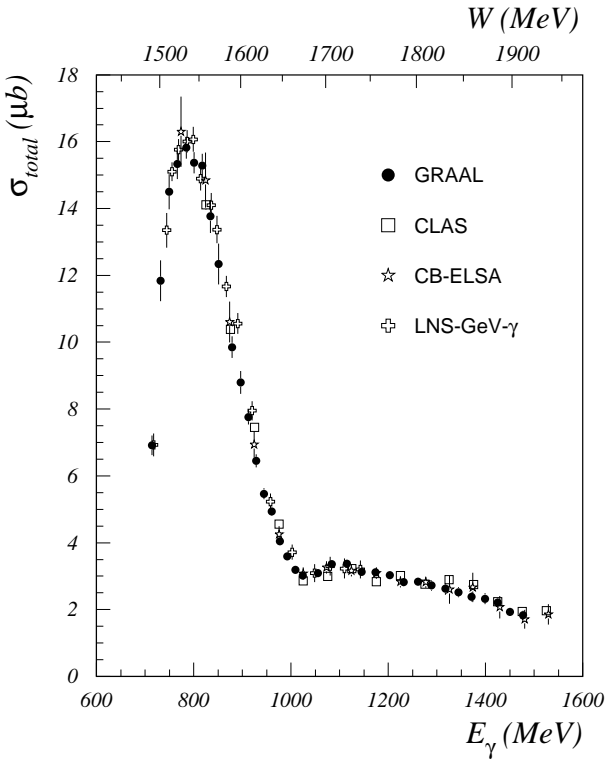} 
\end{center}
\caption{Estimated total cross-section. The GRAAL results (closed
circles) are compared with CLAS (open squares), CB-ELSA (open stars) and LNS-GeV-$\gamma$ (open crosses) data.}
\label{sect_tot}
\end{figure}

The total cross section, plotted in fig. \ref{sect_tot}, was obtained by integration of the 
measured differential cross section,
using the Bonn-Gatchina model (see sect. \ref{disc}) to extrapolate to the uncovered forward 
region (5 to 15\% of the full angular range, depending on the energy). The extrapolated 
fraction amounts between 2 and 15\% 
of the estimated total cross section for most of the energies.
The plotted errors were calculated from the experimental ones and include an additional
uncertainty due to the extrapolation procedure. The latter was estimated from the variation of
the total cross section when considering for extrapolation the two other models discussed below,
whose behaviours in the most forward region differ from the Bonn-Gatchina model
(see figs. \ref{sect1} and \ref{sect2}). The resulting error represents at the most 5\% 
of the total cross section. 

It should be noted that, despite a good agreement between the differential cross sections, 
the new total cross section is significantly lower than our previous estimate
in the 1050-1100 MeV range. This discrepancy mostly originates from the forward
region. First, as already stated, the new data are much more precise in this region
and clearly indicate a drop of the cross section at forward angles, not seen before.
The integral over the measured range is now lower. Second, in agreement with our data,
the model presently used to extrapolate drops at forward angles; it
gives therefore a smaller contribution as compared to the simple polynomial fit (degree two)
used in our previous publication. Both effects do explain the discrepancy.

Fig. \ref{sect_tot} displays the comparison with the CLAS \cite{dug02}, CB-ELSA \cite{cre05} 
and LNS-GeV-$\gamma$ \cite{nak06} results. Apart from LNS-GeV-$\gamma$, CLAS and CB-ELSA also
did not cover the full angular range and had to extrapolate to estimate total cross section.
They used MAID and Bonn-Gatchina models, respectively. The new GRAAL estimates agree now well 
with all other results.

\subsection{Discussion}
\label{disc}

We have compared our results with three models:  
the isobar model MAID, the coupled-channel partial-wave analysis developed by the Bonn-Gatchina 
group and the constituent quark model of Saghai and Li. In the following, 
these two latest models will be referred as BCC and CQM, respectively.

The MAID model \cite{chi03} is an isobar model designed to fit the $\eta$ photo-
and electroproduction database. This model contains, besides Born terms and vector meson exchanges,
contributions from the following well-established resonances: $D_{13}$(1520), $S_{11}$(1535), 
$S_{11}$(1650), $D_{15}$(1675), $F_{15}$(1680), $D_{13}$(1700), $P_{11}$(1710) and $P_{13}$(1720). 
The model is fitted to current photoproduction cross section data from Mainz-TAPS \cite{kru95}, 
GRAAL \cite{ren02} and CLAS \cite{dug02} as well as beam asymmetries from GRAAL \cite{aja98}.
The fit gives resonance masses and widths in good agreement with the PDG 
compilation \cite{pdg04}.

In an alternative MAID analysis, the standard treatment of t-channel vector 
meson exchange is replaced by Regge trajectories while keeping the same $N^*$ contributions \cite{chi04}. 
The reggeized version of the MAID model is dedicated to fit the $\eta$ and $\eta'$ photoproduction 
database. Both standard and reggeized models give an overall good description of the current $\eta$ 
photoproduction results in the resonance region (W $\leq$ 2 GeV), the reggeized model becoming more 
appropriate to describe higher energy data. It was however found that 
the standard isobar model leads to an unusually large $\eta N$ branching ratio (17\%)
for the $D_{15}$(1675) resonance, whereas the reggeized model requires a rather small coupling (0.7\%) 
\cite{tia06}. 

The BCC model \cite{ani05,sar05} is a combined analysis of photoproduction experiments with 
$\pi N$, $\eta N$, $K\Lambda$ and $K\Sigma$ final states. 
$\pi^0$ and $\eta$ photoproduction data from CB-ELSA \cite{cre05,bat05}, Mainz-TAPS \cite{kru95} 
and GRAAL \cite{aja98,bar05} as well as results on $\gamma p \rightarrow n\pi^+$ \cite{npi06} were used. 
Data available from SAPHIR \cite{gla04} , CLAS \cite{nab04} and LEPS \cite{zeg03,law05} for the 
reactions $\gamma p \rightarrow K^+\Lambda$, $\gamma p \rightarrow K^+ \Sigma^0$ and 
$\gamma p \rightarrow K^0 \Sigma^+$ were also included. As compared to the other models, the BCC 
partial wave analysis takes into account a much larger database. A fair agreement with the whole 
database was obtained with 14 $N^*$ and 7 $\Delta^*$ resonances whose masses, widths and 
electromagnetic amplitudes are compatible with the PDG compilation \cite{pdg04}. One of the main 
outcome of this model is the necessity to introduce several new resonances above 1800~MeV, in
particular the $D_{13}$(1875) and $D_{15}$(2070) nucleonic states. The $D_{15}$(2070) resonance is 
found to have a sizeable coupling to the $\eta$N final state while the $D_{13}$(1875) 
does not significantly contribute. On the other hand, this latter is found to have larger 
couplings to the K$\Lambda$ and K$\Sigma$ final states as confirmed by our recently published results 
on $K\Lambda$ and $K\Sigma^0$ photoproduction \cite{lle07}.

\begin{figure}
\begin{center}
\includegraphics[width=0.9\linewidth]{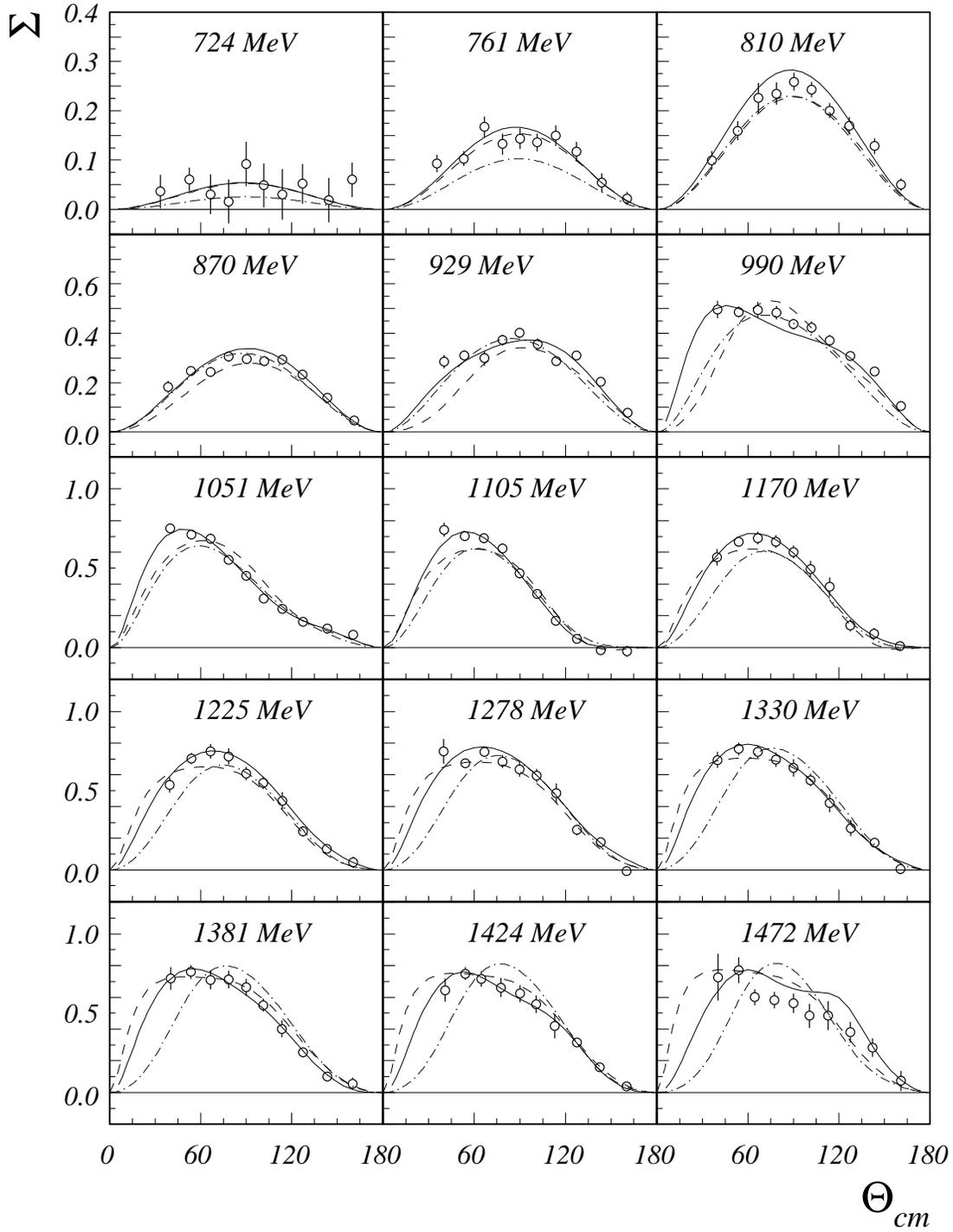} 
\end{center}
\caption{Angular distributions of the beam asymmetry. Data are compared 
with results of the MAID (dashed line), CQM (dotted-dashed line) and BCC (solid line) models.}
\label{ast}
\end{figure}

\begin{figure}
\begin{center}
\includegraphics[width=0.9\linewidth]{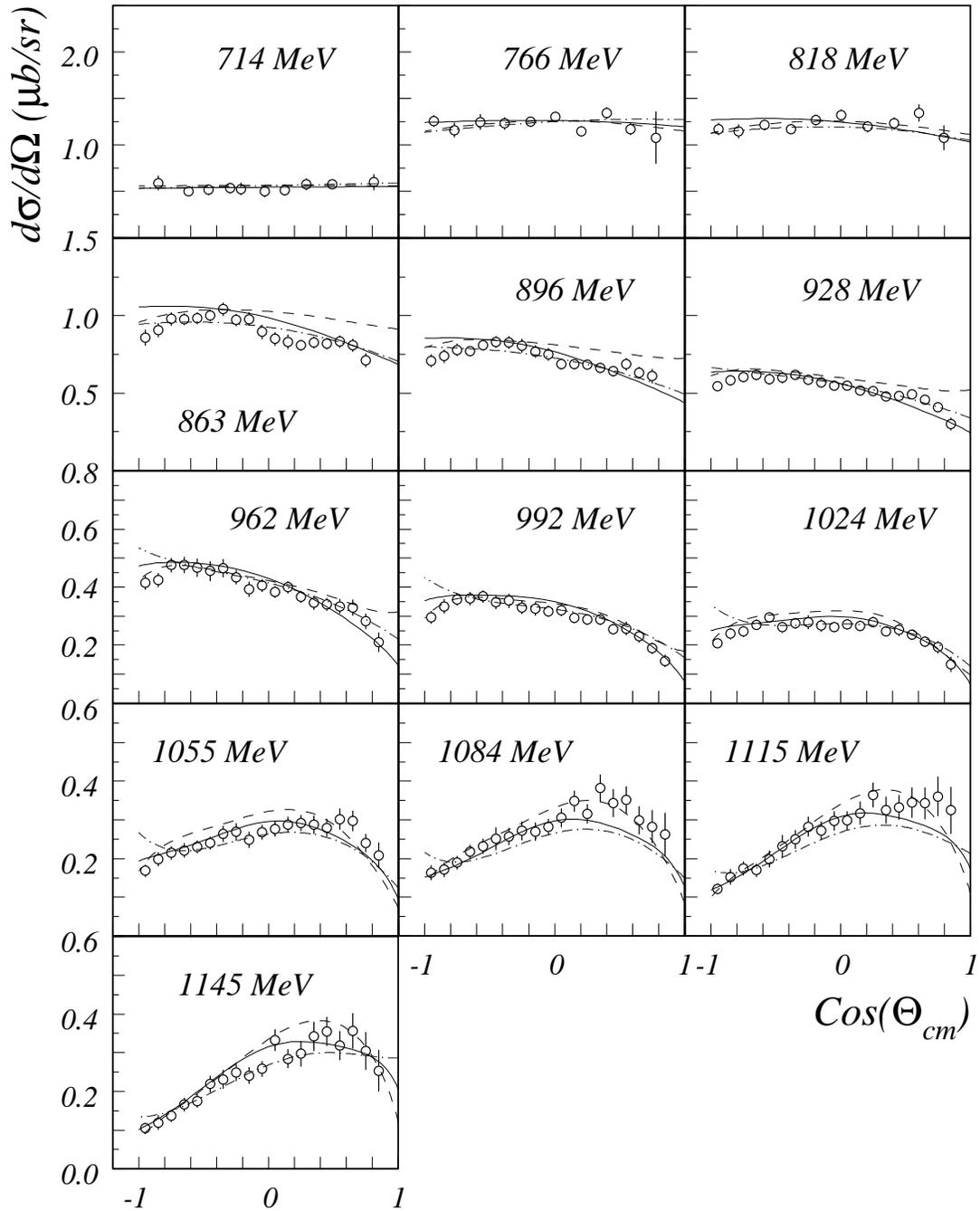} 
\end{center}
\caption{Differential cross section for energies ranging from threshold to 1150 MeV.
Curve definition as in fig. \ref{ast}.}
\label{sect1}
\end{figure}

\begin{figure}
\begin{center}
\includegraphics[width=0.9\linewidth]{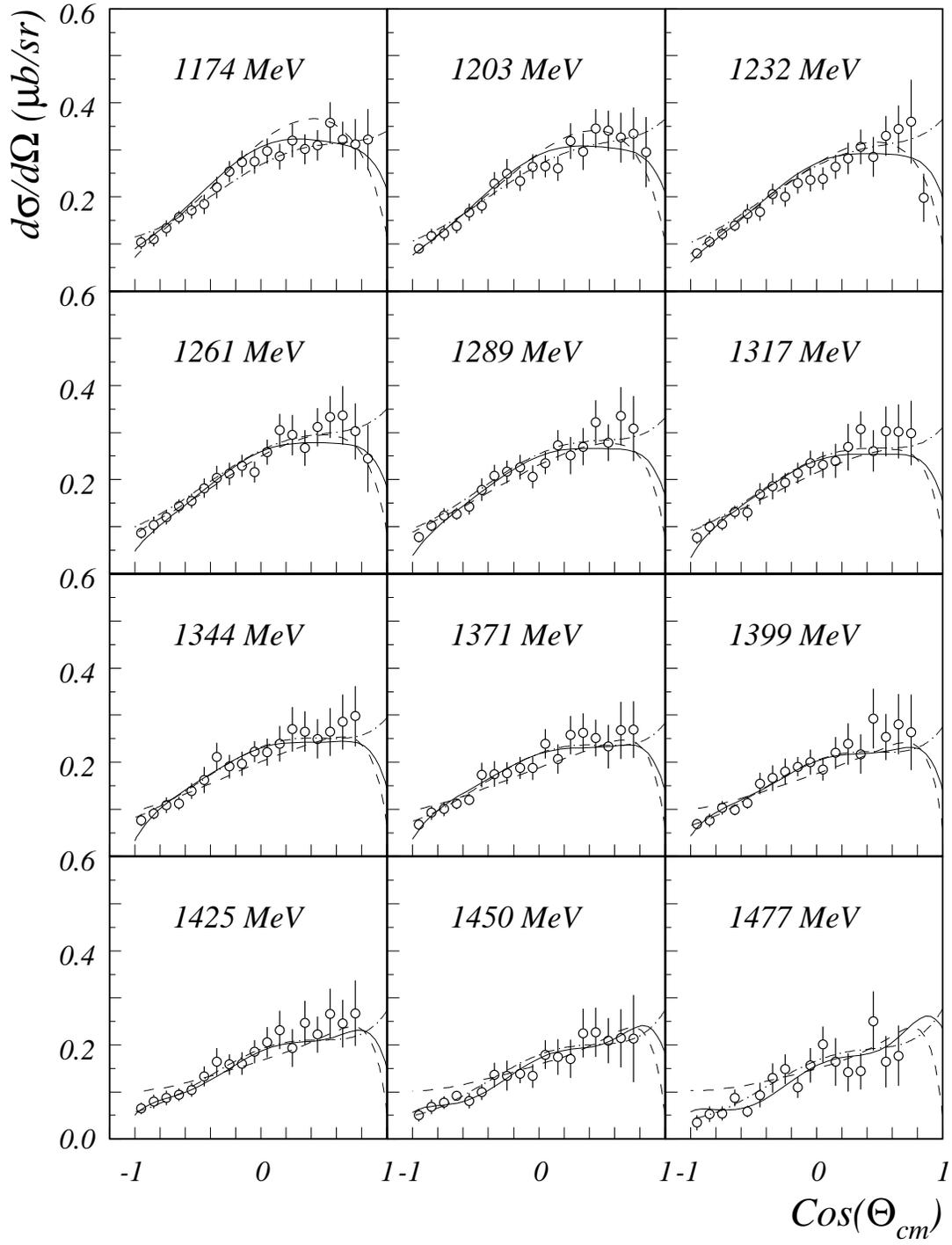} 
\end{center}
\caption{Differential cross section for energies ranging from 1150 to 1500 MeV.
Curve definition as in fig. \ref{ast}.}
\label{sect2}
\end{figure}

The results of the standard MAID \cite{mai06} and BCC \cite{ani05} models presented in 
figs.~\ref{asym_comp} and \ref{ast} to \ref{sect2} include
in their respective database our previously published data up to 1100 MeV and some preliminary beam 
asymmetry values above 1100 MeV. These two models were not re-fitted to take into account our 
final data set. For both observables, the overall agreement with the MAID model (dashed line) 
is quite satisfactory. For the BCC model (solid line), the agreement is
also very good and even better for the beam asymmetry. However, contributions of individual 
resonances other than the dominant
$S_{11}$(1535) (as well as $S_{11}$(1650)) differ for both models. The fit with 
the MAID model indeed requires a strong contribution from the $P_{11}$(1710) partial wave whereas the 
BCC model needs a strong $P_{13}$(1720) state (the $P_{11}$(1710) plays no role).
By contrast, the $D_{15}$(1675) resonance is negligible in the BCC analysis while it 
has a sizeable contribution in the MAID model.

The CQM model \cite{li98,sag01} is a chiral constituent quark model and embodies all known nucleonic
resonances (the same as in MAID plus $P_{13}$(1900) and $F_{15}$(2000)).
The fitted database contains differential cross sections and beam asymmetries from Mainz-TAPS \cite{kru95}, 
GRAAL \cite{aja98,ren02}, CLAS \cite{dug02} and CB-ELSA \cite{cre05,bat05} up to 2 GeV. 
Despite the presence of all known resonances, the model could not fit properly our previously published set
of data. Only the introduction of a new $S_{11}$ resonance
allowed to reproduce nicely the experimental data \cite{ren02}. According to the authors, this resonance, 
not predicted by the Constituent Quark Model, may have an exotic nature such as a $\Sigma K$ or 
$\Lambda K$ molecular state. 

The inclusion of our new data in the CQM model has started only recently and the conclusions are
still preliminary. Nevertheless, the new fit (dotted-dashed line in figs.
\ref{ast} to \ref{sect2}) confirms the necessity of a third $S_{11}$ resonance with a mass of 
1730 MeV and a width of 240~MeV \cite{sag06}. It needs also the presence of the two new resonances $D_{13}$(1875) 
and $D_{15}$(2070) with masses and widths in agreement with the predictions of the Bonn-Gatchina 
model.

\begin{figure}
\begin{center}
\includegraphics[width=1.1\linewidth]{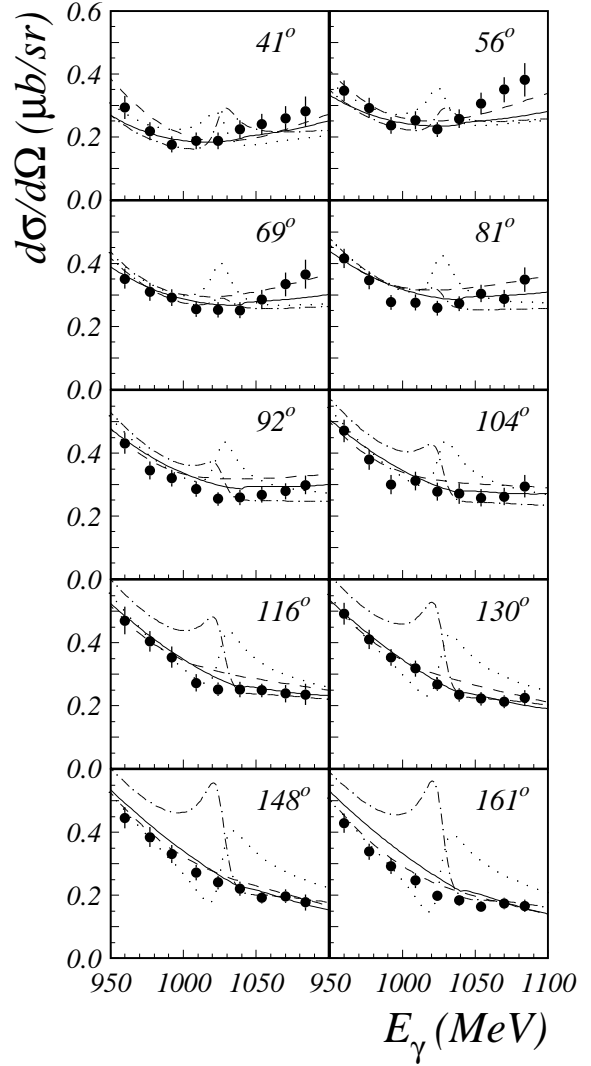} 
\end{center}
\caption{Differential cross section between 950 and 1100 MeV obtained with a narrow energy binning
for various $\eta$ center-of-mass angles.
Comparison with the standard MAID model (dashed line), BCC partial-wave analysis (solid line) and predictions
of the modified reggeized MAID model including a narrow $P_{11}$ state. For this latter model, two versions
are displayed corresponding to the two choices for the $\zeta_{\eta N}$ hadronic relative phase (dotted-dashed line: 
$\zeta_{\eta N}=+1$ - dotted line: $\zeta_{\eta N}=-1$).}
\label{e_fin1}
\end{figure}

\begin{figure}
\begin{center}
\includegraphics[width=1.1\linewidth]{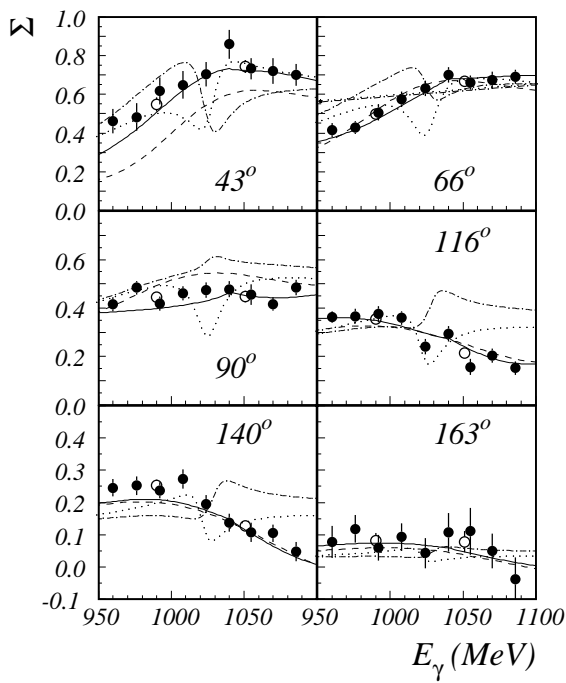} 
\end{center}
\caption{Beam asymmetry between 950 and 1100 MeV obtained with a narrow energy binning ($\sim$16~MeV) for various 
$\eta$ center-of-mass angles (closed circles). Data corresponding to the large energy binning ($\sim$60~MeV) 
presented in fig.~\ref{asym_comp} are also plotted (open circles). Curve definition as in fig.~\ref{e_fin1}.}
\label{e_fin2}
\end{figure}

Recent experimental and theoretical works in connection to the search for narrow exotic states
have focused the attention on $\eta$ photoproduction on both proton and neutron in the energy 
region around 1~GeV. Preliminary cross section data on quasi-free neutron have been recently obtained 
by the GRAAL \cite{reb05}, CB-ELSA/TAPS \cite{jae05} and LNS-GeV-$\gamma$ \cite{kas06} collaborations. 
These results exhibit, in addition to the dominant $S_{11}$(1535), a resonant structure around 
E$_\gamma$=1 GeV (W=1.67~GeV), not seen on the proton to date. Beam asymmetries on quasi-free neutron 
have also been measured by our collaboration and will be presented and discussed in a forthcoming article \cite{fan07}.
Several theoretical works have been recently performed to provide an explanation of the structure
seen in the cross section in terms of a baryon resonance predominantly coupled to the neutron. 
In the framework of the standard MAID model, this bump could be assigned to the $D_{15}$(1675) 
resonance \cite{tia06}. The coupled-channel Giessen model shows that this peak could 
be interpreted by the $S_{11}$(1650) 
and $P_{11}$(1710) excitations \cite{shk06}. By contrast, a modified version of the reggeized MAID model 
shows that the inclusion of an additional exotic narrow $P_{11}$(1670) state, with a width of 10-30 MeV, 
could explain the observed structure \cite{fix07}. This resonance was suggested in some previous works
\cite{pol03}-\cite{dia04} to be the nucleon-like member of the anti-decuplet of pentaquarks predicted by the 
chiral soliton model \cite{dia97}.
The modified reggeized MAID calculation predicts that this state, although much less coupled to
the proton, should also be visible in $\eta$ photoproduction on the proton. A 
pronounced narrow structure should be seen mainly at backward angles in the differential cross 
section and at all angles in the beam asymmetry \cite{fix07}.

In order to look for this narrow structure, we extracted the differential cross section and beam 
asymmetry with the finest energy binning compatible with the energy resolution ($\Delta E_\gamma \sim$16~MeV). 
The results obtained between 950 and 1100 MeV are presented in figs. \ref{e_fin1} and \ref{e_fin2}
for various $\eta$ center-of-mass angles ranging from 40$^0$ to 160$^0$. 
Neither the differential cross section nor the beam asymmetry do show any evidence of a narrow structure.
From the differential cross section and beam asymmetry results extracted for each of the different
data taking periods, it was checked that no robust narrow signal was hidden or smeared by the data merging.
The standard MAID and BCC models remain in fair agreement with our data even with the finer energy binning.
In addition, the predictions \cite{fix07},\cite{tia07} of modified versions of the reggeized MAID model, 
including a narrow $P_{11}$(1670) state (10 MeV width), exhibit structures incompatible with our data.

\section{Summary}

In this paper, we have presented high precision measurements of the differential cross section and beam
asymmetry for the $\gamma$p$\rightarrow\eta$p reaction, from threshold to 1500 MeV. The results
are in good agreement with all previously published data. For this channel, an extensive database 
containing accurate beam asymmetries together with differential cross sections is now available.
Various models are able to nicely fit these results but, despite constraints brought
by the beam asymmetry, their conclusions remain different in terms of individual resonance
contributions. New measurements on other polarization observables are therefore necessary to resolve
these ambiguities. The possible contribution of a narrow state N(1670) was also investigated
and no evidence was found.

\vspace{5mm}

\noindent
{\bf Acknowledgements}

\noindent
We are grateful to A.V. Anisovitch, B. Saghai and L. Tiator for fruitful discussions and communication of their analyses.
We thank J. Kasagi for communication of the LNS cross section data. The support of the technical groups from all contributing
institutions is greatly acknowledged. It is a pleasure to thank the ESRF as a host institution and its technical staff for 
the smooth operation of the storage ring.

\newpage

\onecolumn

\newpage

\onecolumn

\begin{table}
\begin{center}
\caption{Beam asymmetry $\Sigma$ as a function of the photon laboratory energy and the $\eta$ center-of-mass angle.}
\label{table_ass1}

\vspace{0.3cm}

\begin{tabular}{|r|r|r|r|r|r|r|r|} \hline

\multicolumn{1}{|c|}{$\theta_{cm} (^o)$} & \multicolumn{1}{|c|}{$E_{\gamma}$=724 MeV} &
\multicolumn{1}{|c|}{$\theta_{cm} (^o)$} & \multicolumn{1}{|c|}{$E_{\gamma}$=761 MeV} &
\multicolumn{1}{|c|}{$\theta_{cm} (^o)$} & \multicolumn{1}{|c|}{$E_{\gamma}$=810 MeV} &
\multicolumn{1}{|c|}{$\theta_{cm} (^o)$} & \multicolumn{1}{|c|}{$E_{\gamma}$=870 MeV} \\ \hline

  33.4 &  0.036 $\pm$ 0.034 &   35.4 &  0.093 $\pm$ 0.018 &   36.1 &  0.100 $\pm$ 0.018 &   38.7 &  0.182 $\pm$ 0.026 \\ 
  52.6 &  0.061 $\pm$ 0.023 &   53.1 &  0.103 $\pm$ 0.015 &   53.3 &  0.159 $\pm$ 0.019 &   53.6 &  0.246 $\pm$ 0.015 \\ 
  66.4 &  0.031 $\pm$ 0.040 &   66.5 &  0.167 $\pm$ 0.021 &   66.7 &  0.226 $\pm$ 0.030 &   66.5 &  0.243 $\pm$ 0.019 \\
  78.4 &  0.016 $\pm$ 0.045 &   78.5 &  0.132 $\pm$ 0.021 &   78.8 &  0.235 $\pm$ 0.022 &   78.5 &  0.305 $\pm$ 0.020 \\
  90.0 &  0.092 $\pm$ 0.045 &   90.0 &  0.143 $\pm$ 0.021 &   90.3 &  0.259 $\pm$ 0.018 &   90.2 &  0.295 $\pm$ 0.018 \\
 101.6 &  0.049 $\pm$ 0.045 &  101.5 &  0.136 $\pm$ 0.019 &  101.8 &  0.243 $\pm$ 0.016 &  101.8 &  0.288 $\pm$ 0.020 \\
 113.9 &  0.031 $\pm$ 0.051 &  113.6 &  0.150 $\pm$ 0.021 &  113.8 &  0.200 $\pm$ 0.015 &  114.0 &  0.293 $\pm$ 0.020 \\
 127.1 &  0.052 $\pm$ 0.040 &  127.1 &  0.117 $\pm$ 0.020 &  126.9 &  0.170 $\pm$ 0.017 &  127.0 &  0.233 $\pm$ 0.012 \\
 144.5 &  0.019 $\pm$ 0.045 &  143.9 &  0.054 $\pm$ 0.019 &  143.6 &  0.128 $\pm$ 0.016 &  143.4 &  0.138 $\pm$ 0.013 \\
 159.8 &  0.060 $\pm$ 0.035 &  160.8 &  0.022 $\pm$ 0.013 &  161.0 &  0.050 $\pm$ 0.013 &  161.1 &  0.047 $\pm$ 0.010 \\ \hline

\multicolumn{1}{|c|}{$\theta_{cm} (^o)$} & \multicolumn{1}{|c|}{$E_{\gamma}$=929 MeV}  &
\multicolumn{1}{|c|}{$\theta_{cm} (^o)$} & \multicolumn{1}{|c|}{$E_{\gamma}$=990 MeV}  &
\multicolumn{1}{|c|}{$\theta_{cm} (^o)$} & \multicolumn{1}{|c|}{$E_{\gamma}$=1051 MeV} &
\multicolumn{1}{|c|}{$\theta_{cm} (^o)$} & \multicolumn{1}{|c|}{$E_{\gamma}$=1105 MeV} \\ \hline

  40.1 &  0.285 $\pm$ 0.028 &	40.1 &  0.497 $\pm$ 0.036 &   39.9 &  0.750 $\pm$ 0.032 &   40.1 &  0.742 $\pm$ 0.045 \\
  53.5 &  0.310 $\pm$ 0.021 &   53.7 &  0.486 $\pm$ 0.022 &   53.8 &  0.713 $\pm$ 0.022 &   53.8 &  0.702 $\pm$ 0.026 \\  
  66.6 &  0.299 $\pm$ 0.030 &   66.6 &  0.494 $\pm$ 0.033 &   66.5 &  0.686 $\pm$ 0.033 &   66.5 &  0.687 $\pm$ 0.032 \\
  78.5 &  0.372 $\pm$ 0.022 &   78.6 &  0.484 $\pm$ 0.028 &   78.6 &  0.553 $\pm$ 0.025 &   78.5 &  0.623 $\pm$ 0.032 \\
  89.9 &  0.402 $\pm$ 0.021 &   90.0 &  0.437 $\pm$ 0.022 &   89.9 &  0.450 $\pm$ 0.028 &   90.0 &  0.467 $\pm$ 0.031 \\
 101.9 &  0.355 $\pm$ 0.024 &  101.7 &  0.424 $\pm$ 0.025 &  101.5 &  0.306 $\pm$ 0.026 &  101.3 &  0.338 $\pm$ 0.028 \\
 113.9 &  0.287 $\pm$ 0.019 &  113.8 &  0.371 $\pm$ 0.023 &  113.9 &  0.242 $\pm$ 0.026 &  113.8 &  0.168 $\pm$ 0.028 \\
 127.3 &  0.310 $\pm$ 0.017 &  127.4 &  0.307 $\pm$ 0.018 &  127.4 &  0.162 $\pm$ 0.023 &  127.5 &  0.053 $\pm$ 0.023 \\
 143.4 &  0.203 $\pm$ 0.015 &  143.5 &  0.244 $\pm$ 0.018 &  143.4 &  0.120 $\pm$ 0.018 &  143.2 & -0.019 $\pm$ 0.022 \\
 160.9 &  0.077 $\pm$ 0.014 &  160.7 &  0.104 $\pm$ 0.019 &  160.6 &  0.080 $\pm$ 0.019 &  160.5 & -0.026 $\pm$ 0.033 \\ \hline

\multicolumn{1}{|c|}{$\theta_{cm} (^o)$} & \multicolumn{1}{|c|}{$E_{\gamma}$=1170 MeV} &
\multicolumn{1}{|c|}{$\theta_{cm} (^o)$} & \multicolumn{1}{|c|}{$E_{\gamma}$=1225 MeV} &
\multicolumn{1}{|c|}{$\theta_{cm} (^o)$} & \multicolumn{1}{|c|}{$E_{\gamma}$=1278 MeV} &
\multicolumn{1}{|c|}{$\theta_{cm} (^o)$} & \multicolumn{1}{|c|}{$E_{\gamma}$=1330 MeV} \\ \hline

  39.6 &  0.568 $\pm$ 0.051 &	39.7 &  0.536 $\pm$ 0.048 &   39.9 &  0.748 $\pm$ 0.077 &   40.0 &  0.694 $\pm$ 0.051 \\
  53.8 &  0.667 $\pm$ 0.031 &   53.8 &  0.701 $\pm$ 0.036 &   53.9 &  0.671 $\pm$ 0.029 &   53.7 &  0.763 $\pm$ 0.042  \\
  66.5 &  0.690 $\pm$ 0.041 &   66.6 &  0.748 $\pm$ 0.045 &   66.5 &  0.745 $\pm$ 0.031 &   66.6 &  0.744 $\pm$ 0.039 \\
  78.5 &  0.668 $\pm$ 0.040 &   78.5 &  0.716 $\pm$ 0.051 &   78.5 &  0.684 $\pm$ 0.040 &   78.6 &  0.697 $\pm$ 0.045 \\
  89.9 &  0.601 $\pm$ 0.041 &   90.0 &  0.608 $\pm$ 0.041 &   89.9 &  0.634 $\pm$ 0.046 &   90.1 &  0.645 $\pm$ 0.055 \\
 101.3 &  0.493 $\pm$ 0.051 &  101.2 &  0.549 $\pm$ 0.038 &  101.2 &  0.594 $\pm$ 0.043 &  101.3 &  0.567 $\pm$ 0.041 \\
 113.8 &  0.382 $\pm$ 0.061 &  113.8 &  0.437 $\pm$ 0.054 &  113.9 &  0.482 $\pm$ 0.064 &  113.8 &  0.422 $\pm$ 0.057 \\
 127.4 &  0.138 $\pm$ 0.040 &  127.5 &  0.243 $\pm$ 0.032 &  127.6 &  0.252 $\pm$ 0.033 &  127.7 &  0.264 $\pm$ 0.050 \\
 143.1 &  0.085 $\pm$ 0.036 &  143.2 &  0.132 $\pm$ 0.028 &  143.2 &  0.176 $\pm$ 0.028 &  143.3 &  0.172 $\pm$ 0.028 \\
 160.2 &  0.009 $\pm$ 0.030 &  160.4 &  0.048 $\pm$ 0.035 &  160.4 & -0.009 $\pm$ 0.028 &  160.3 &  0.006 $\pm$ 0.033 \\ \hline
 
\end{tabular}

\begin{tabular}{|r|r|r|r|r|r|} \hline
 
\multicolumn{1}{|c|}{$\theta_{cm} (^o)$} & \multicolumn{1}{|c|}{$E_{\gamma}$=1381 MeV} &
\multicolumn{1}{|c|}{$\theta_{cm} (^o)$} & \multicolumn{1}{|c|}{$E_{\gamma}$=1424 MeV} &
\multicolumn{1}{|c|}{$\theta_{cm} (^o)$} & \multicolumn{1}{|c|}{$E_{\gamma}$=1472 MeV} \\ \hline

  40.2 &  0.720 $\pm$ 0.072 &   40.8 &  0.646 $\pm$ 0.073 &   40.2 &  0.727 $\pm$ 0.147 \\ 
  53.5 &  0.760 $\pm$ 0.045 &   54.1 &  0.750 $\pm$ 0.041 &   53.8 &  0.771 $\pm$ 0.080 \\
  66.5 &  0.710 $\pm$ 0.059 &   64.6 &  0.716 $\pm$ 0.048 &   64.6 &  0.601 $\pm$ 0.051 \\
  78.5 &  0.712 $\pm$ 0.056 &   77.5 &  0.661 $\pm$ 0.060 &   77.4 &  0.581 $\pm$ 0.052 \\
  89.9 &  0.663 $\pm$ 0.053 &   90.0 &  0.623 $\pm$ 0.054 &   90.0 &  0.563 $\pm$ 0.062 \\
 101.1 &  0.551 $\pm$ 0.042 &  100.6 &  0.556 $\pm$ 0.055 &  100.6 &  0.483 $\pm$ 0.076 \\
 113.7 &  0.401 $\pm$ 0.053 &  113.0 &  0.419 $\pm$ 0.076 &  113.0 &  0.484 $\pm$ 0.091 \\
 127.6 &  0.252 $\pm$ 0.032 &  127.5 &  0.314 $\pm$ 0.028 &  127.6 &  0.380 $\pm$ 0.064 \\
 143.3 &  0.101 $\pm$ 0.028 &  142.5 &  0.160 $\pm$ 0.023 &  142.2 &  0.282 $\pm$ 0.059 \\
 160.2 &  0.056 $\pm$ 0.038 &  160.3 &  0.038 $\pm$ 0.027 &  160.8 &  0.074 $\pm$ 0.063 \\ \hline
\end{tabular}

\end{center}
\end{table}

\newpage

\onecolumn

\begin{table}
\begin{center}
\caption{Differential cross section $d\sigma/d\Omega$ ($\mu b/sr$) as a function of the photon laboratory energy (700-850 MeV) 
and the cosine of the $\eta$ center-of-mass angle.}
\label{table_sect0}

\vspace{0.3cm}

\begin{tabular}{|r|r|r|r|r|r|r|r|} \hline

\multicolumn{1}{|c|}{$\cos(\theta_{cm})$} & \multicolumn{1}{|c|}{$E_{\gamma}$=714 MeV} &
\multicolumn{1}{|c|}{$\cos(\theta_{cm})$} & \multicolumn{1}{|c|}{$E_{\gamma}$=732 MeV} &
\multicolumn{1}{|c|}{$\cos(\theta_{cm})$} & \multicolumn{1}{|c|}{$E_{\gamma}$=749 MeV} &
\multicolumn{1}{|c|}{$\cos(\theta_{cm})$} & \multicolumn{1}{|c|}{$E_{\gamma}$=766 MeV} \\ \hline

 -0.85 &  0.588 $\pm$  0.077 &  -0.91 &  1.007 $\pm$  0.064 &  -0.92 &  1.108 $\pm$  0.060 &  -0.93 &  1.253 $\pm$  0.065 \\
 -0.62 &  0.499 $\pm$  0.048 &  -0.74 &  0.903 $\pm$  0.060 &  -0.76 &  1.102 $\pm$  0.069 &  -0.77 &  1.158 $\pm$  0.077 \\
 -0.46 &  0.517 $\pm$  0.057 &  -0.54 &  0.884 $\pm$  0.074 &  -0.56 &  1.180 $\pm$  0.089 &  -0.57 &  1.244 $\pm$  0.085 \\
 -0.30 &  0.534 $\pm$  0.059 &  -0.35 &  0.921 $\pm$  0.090 &  -0.38 &  1.246 $\pm$  0.083 &  -0.38 &  1.230 $\pm$  0.068 \\
 -0.21 &  0.523 $\pm$  0.067 &  -0.20 &  0.927 $\pm$  0.094 &  -0.20 &  1.181 $\pm$  0.070 &  -0.19 &  1.246 $\pm$  0.053 \\
 -0.03 &  0.499 $\pm$  0.067 &  -0.02 &  0.839 $\pm$  0.073 &   0.00 &  1.169 $\pm$  0.068 &   0.01 &  1.301 $\pm$  0.053 \\
  0.12 &  0.512 $\pm$  0.060 &   0.17 &  0.863 $\pm$  0.082 &   0.19 &  1.135 $\pm$  0.063 &   0.21 &  1.143 $\pm$  0.055 \\
  0.29 &  0.575 $\pm$  0.065 &   0.38 &  0.985 $\pm$  0.085 &   0.38 &  1.231 $\pm$  0.074 &   0.40 &  1.342 $\pm$  0.062 \\
  0.49 &  0.577 $\pm$  0.050 &   0.59 &  0.928 $\pm$  0.292 &   0.58 &  1.242 $\pm$  0.214 &   0.58 &  1.170 $\pm$  0.069 \\
  0.81 &  0.596 $\pm$  0.088 &   0.81 &  0.943 $\pm$  0.784 &   0.79 &  0.876 $\pm$  0.637 &   0.78 &  1.077 $\pm$  0.281 \\ \hline

\multicolumn{1}{|c|}{$\cos(\theta_{cm})$} & \multicolumn{1}{|c|}{$E_{\gamma}$=785 MeV} &
\multicolumn{1}{|c|}{$\cos(\theta_{cm})$} & \multicolumn{1}{|c|}{$E_{\gamma}$=801 MeV} &
\multicolumn{1}{|c|}{$\cos(\theta_{cm})$} & \multicolumn{1}{|c|}{$E_{\gamma}$=818 MeV} &
\multicolumn{1}{|c|}{$\cos(\theta_{cm})$} & \multicolumn{1}{|c|}{$E_{\gamma}$=835 MeV} \\ \hline

 -0.94 &  1.220 $\pm$  0.073 &  -0.94 &  1.228 $\pm$  0.067 &  -0.94 &  1.170 $\pm$  0.067 &  -0.94 &  1.007 $\pm$  0.066 \\
 -0.78 &  1.289 $\pm$  0.086 &  -0.79 &  1.213 $\pm$  0.081 &  -0.79 &  1.144 $\pm$  0.076 &  -0.79 &  1.082 $\pm$  0.072 \\
 -0.58 &  1.240 $\pm$  0.079 &  -0.58 &  1.245 $\pm$  0.065 &  -0.59 &  1.214 $\pm$  0.058 &  -0.59 &  1.117 $\pm$  0.052 \\
 -0.39 &  1.286 $\pm$  0.060 &  -0.38 &  1.230 $\pm$  0.050 &  -0.38 &  1.165 $\pm$  0.047 &  -0.38 &  1.167 $\pm$  0.048 \\
 -0.19 &  1.281 $\pm$  0.056 &  -0.19 &  1.257 $\pm$  0.054 &  -0.19 &  1.268 $\pm$  0.059 &  -0.19 &  1.172 $\pm$  0.053 \\
  0.01 &  1.255 $\pm$  0.056 &   0.01 &  1.251 $\pm$  0.057 &   0.00 &  1.320 $\pm$  0.062 &   0.01 &  1.121 $\pm$  0.059 \\
  0.21 &  1.305 $\pm$  0.067 &   0.20 &  1.324 $\pm$  0.064 &   0.21 &  1.194 $\pm$  0.062 &   0.22 &  1.117 $\pm$  0.060 \\
  0.41 &  1.300 $\pm$  0.066 &   0.40 &  1.160 $\pm$  0.071 &   0.41 &  1.231 $\pm$  0.067 &   0.41 &  1.109 $\pm$  0.059 \\
  0.60 &  1.212 $\pm$  0.071 &   0.60 &  1.186 $\pm$  0.079 &   0.60 &  1.341 $\pm$  0.092 &   0.61 &  1.097 $\pm$  0.082 \\
  0.79 &  1.287 $\pm$  0.130 &   0.79 &  1.181 $\pm$  0.097 &   0.79 &  1.075 $\pm$  0.134 &   0.80 &  1.046 $\pm$  0.393 \\ \hline

\end{tabular}

\end{center}
\end{table}

\newpage

\onecolumn

\begin{table}
\begin{center}
\caption{Differential cross section $d\sigma/d\Omega$ ($\mu b/sr$) as a function of the photon laboratory energy (850-1200 MeV) 
and the cosine of the $\eta$ center-of-mass angle.}
\label{table_sect1}

\vspace{0.3cm}

\begin{tabular}{|r|r|r|r|r|r|r|r|} \hline

\multicolumn{1}{|c|}{$\cos(\theta_{cm})$} & \multicolumn{1}{|c|}{$E_{\gamma}$=863 MeV}  &
\multicolumn{1}{|c|}{$\cos(\theta_{cm})$} & \multicolumn{1}{|c|}{$E_{\gamma}$=896 MeV}  &
\multicolumn{1}{|c|}{$\cos(\theta_{cm})$} & \multicolumn{1}{|c|}{$E_{\gamma}$=928 MeV} &
\multicolumn{1}{|c|}{$\cos(\theta_{cm})$} & \multicolumn{1}{|c|}{$E_{\gamma}$=962 MeV} \\ \hline

 -0.95 &  0.858 $\pm$  0.052 &  -0.95 &  0.707 $\pm$  0.042 &  -0.95 &  0.542 $\pm$  0.030 &  -0.95 &  0.414 $\pm$  0.024 \\
 -0.85 &  0.906 $\pm$  0.045 &  -0.85 &  0.738 $\pm$  0.044 &  -0.85 &  0.582 $\pm$  0.033 &  -0.85 &  0.424 $\pm$  0.025 \\
 -0.75 &  0.980 $\pm$  0.044 &  -0.75 &  0.778 $\pm$  0.041 &  -0.75 &  0.604 $\pm$  0.031 &  -0.75 &  0.475 $\pm$  0.025 \\
 -0.65 &  0.976 $\pm$  0.038 &  -0.65 &  0.771 $\pm$  0.032 &  -0.65 &  0.615 $\pm$  0.032 &  -0.65 &  0.476 $\pm$  0.027 \\
 -0.55 &  0.984 $\pm$  0.038 &  -0.55 &  0.808 $\pm$  0.035 &  -0.55 &  0.589 $\pm$  0.034 &  -0.55 &  0.467 $\pm$  0.032 \\
 -0.45 &  1.001 $\pm$  0.038 &  -0.45 &  0.831 $\pm$  0.038 &  -0.45 &  0.599 $\pm$  0.039 &  -0.45 &  0.455 $\pm$  0.035 \\
 -0.35 &  1.042 $\pm$  0.044 &  -0.35 &  0.825 $\pm$  0.042 &  -0.35 &  0.616 $\pm$  0.031 &  -0.35 &  0.465 $\pm$  0.032 \\
 -0.25 &  0.971 $\pm$  0.040 &  -0.25 &  0.805 $\pm$  0.044 &  -0.25 &  0.584 $\pm$  0.034 &  -0.25 &  0.434 $\pm$  0.026 \\
 -0.15 &  0.978 $\pm$  0.043 &  -0.15 &  0.769 $\pm$  0.040 &  -0.15 &  0.568 $\pm$  0.036 &  -0.15 &  0.392 $\pm$  0.026 \\
 -0.05 &  0.895 $\pm$  0.047 &  -0.05 &  0.749 $\pm$  0.042 &  -0.05 &  0.546 $\pm$  0.032 &  -0.05 &  0.406 $\pm$  0.023 \\
  0.05 &  0.849 $\pm$  0.043 &   0.05 &  0.686 $\pm$  0.036 &   0.05 &  0.548 $\pm$  0.027 &   0.05 &  0.384 $\pm$  0.020 \\
  0.15 &  0.831 $\pm$  0.047 &   0.15 &  0.685 $\pm$  0.033 &   0.15 &  0.516 $\pm$  0.027 &   0.15 &  0.399 $\pm$  0.022 \\
  0.25 &  0.809 $\pm$  0.036 &   0.25 &  0.683 $\pm$  0.033 &   0.25 &  0.514 $\pm$  0.027 &   0.25 &  0.366 $\pm$  0.021 \\
  0.35 &  0.827 $\pm$  0.038 &   0.35 &  0.661 $\pm$  0.032 &   0.35 &  0.479 $\pm$  0.026 &   0.35 &  0.346 $\pm$  0.023 \\
  0.45 &  0.818 $\pm$  0.038 &   0.45 &  0.640 $\pm$  0.033 &   0.45 &  0.480 $\pm$  0.028 &   0.45 &  0.340 $\pm$  0.022 \\
  0.55 &  0.832 $\pm$  0.041 &   0.55 &  0.686 $\pm$  0.040 &   0.55 &  0.490 $\pm$  0.030 &   0.55 &  0.334 $\pm$  0.025 \\
  0.65 &  0.809 $\pm$  0.043 &   0.65 &  0.632 $\pm$  0.039 &   0.65 &  0.458 $\pm$  0.030 &   0.65 &  0.329 $\pm$  0.028 \\
  0.75 &  0.711 $\pm$  0.045 &   0.75 &  0.609 $\pm$  0.046 &   0.75 &  0.407 $\pm$  0.032 &   0.75 &  0.282 $\pm$  0.027 \\
       &                    &        &                    &   0.85 &  0.300 $\pm$  0.040 &   0.85 &  0.210 $\pm$  0.033 \\ \hline

\multicolumn{1}{|c|}{$\cos(\theta_{cm})$} & \multicolumn{1}{|c|}{$E_{\gamma}$=992 MeV} &
\multicolumn{1}{|c|}{$\cos(\theta_{cm})$} & \multicolumn{1}{|c|}{$E_{\gamma}$=1024 MeV} &
\multicolumn{1}{|c|}{$\cos(\theta_{cm})$} & \multicolumn{1}{|c|}{$E_{\gamma}$=1055 MeV} &
\multicolumn{1}{|c|}{$\cos(\theta_{cm})$} & \multicolumn{1}{|c|}{$E_{\gamma}$=1084 MeV} \\ \hline

 -0.95 &  0.296 $\pm$  0.022 &  -0.95 &  0.206 $\pm$  0.016 &  -0.95 &  0.168 $\pm$  0.016 &  -0.95 &  0.164 $\pm$  0.020 \\
 -0.85 &  0.334 $\pm$  0.025 &  -0.85 &  0.239 $\pm$  0.019 &  -0.85 &  0.199 $\pm$  0.018 &  -0.85 &  0.171 $\pm$  0.019 \\
 -0.75 &  0.358 $\pm$  0.022 &  -0.75 &  0.248 $\pm$  0.018 &  -0.75 &  0.214 $\pm$  0.017 &  -0.75 &  0.190 $\pm$  0.018 \\
 -0.65 &  0.360 $\pm$  0.023 &  -0.65 &  0.270 $\pm$  0.018 &  -0.65 &  0.221 $\pm$  0.019 &  -0.65 &  0.218 $\pm$  0.019 \\
 -0.55 &  0.368 $\pm$  0.021 &  -0.55 &  0.295 $\pm$  0.018 &  -0.55 &  0.230 $\pm$  0.017 &  -0.55 &  0.232 $\pm$  0.020 \\
 -0.45 &  0.348 $\pm$  0.025 &  -0.45 &  0.262 $\pm$  0.019 &  -0.45 &  0.241 $\pm$  0.017 &  -0.45 &  0.252 $\pm$  0.029 \\
 -0.35 &  0.353 $\pm$  0.025 &  -0.35 &  0.275 $\pm$  0.018 &  -0.35 &  0.264 $\pm$  0.021 &  -0.35 &  0.257 $\pm$  0.022 \\
 -0.25 &  0.330 $\pm$  0.024 &  -0.25 &  0.279 $\pm$  0.023 &  -0.25 &  0.269 $\pm$  0.023 &  -0.25 &  0.272 $\pm$  0.024 \\
 -0.15 &  0.325 $\pm$  0.023 &  -0.15 &  0.268 $\pm$  0.021 &  -0.15 &  0.249 $\pm$  0.021 &  -0.15 &  0.269 $\pm$  0.024 \\
 -0.05 &  0.317 $\pm$  0.019 &  -0.05 &  0.262 $\pm$  0.018 &  -0.05 &  0.268 $\pm$  0.020 &  -0.05 &  0.282 $\pm$  0.021 \\
  0.05 &  0.319 $\pm$  0.019 &   0.05 &  0.272 $\pm$  0.018 &   0.05 &  0.276 $\pm$  0.019 &   0.05 &  0.305 $\pm$  0.024 \\
  0.15 &  0.293 $\pm$  0.018 &   0.15 &  0.266 $\pm$  0.018 &   0.15 &  0.288 $\pm$  0.020 &   0.15 &  0.349 $\pm$  0.027 \\
  0.25 &  0.288 $\pm$  0.019 &   0.25 &  0.280 $\pm$  0.017 &   0.25 &  0.292 $\pm$  0.021 &   0.25 &  0.315 $\pm$  0.025 \\
  0.35 &  0.288 $\pm$  0.018 &   0.35 &  0.248 $\pm$  0.017 &   0.35 &  0.287 $\pm$  0.023 &   0.35 &  0.382 $\pm$  0.035 \\
  0.45 &  0.256 $\pm$  0.018 &   0.45 &  0.253 $\pm$  0.022 &   0.45 &  0.280 $\pm$  0.023 &   0.45 &  0.343 $\pm$  0.036 \\
  0.55 &  0.258 $\pm$  0.023 &   0.55 &  0.235 $\pm$  0.018 &   0.55 &  0.302 $\pm$  0.027 &   0.55 &  0.351 $\pm$  0.035 \\
  0.65 &  0.231 $\pm$  0.022 &   0.65 &  0.212 $\pm$  0.019 &   0.65 &  0.297 $\pm$  0.027 &   0.65 &  0.299 $\pm$  0.036 \\
  0.75 &  0.189 $\pm$  0.019 &   0.75 &  0.193 $\pm$  0.021 &   0.75 &  0.240 $\pm$  0.024 &   0.75 &  0.282 $\pm$  0.042 \\
  0.85 &  0.146 $\pm$  0.022 &   0.85 &  0.134 $\pm$  0.026 &   0.85 &  0.208 $\pm$  0.033 &   0.85 &  0.262 $\pm$  0.056 \\ \hline

\multicolumn{1}{|c|}{$\cos(\theta_{cm})$} & \multicolumn{1}{|c|}{$E_{\gamma}$=1115 MeV} &
\multicolumn{1}{|c|}{$\cos(\theta_{cm})$} & \multicolumn{1}{|c|}{$E_{\gamma}$=1145 MeV} &
\multicolumn{1}{|c|}{$\cos(\theta_{cm})$} & \multicolumn{1}{|c|}{$E_{\gamma}$=1174 MeV} &
\multicolumn{1}{|c|}{$\cos(\theta_{cm})$} & \multicolumn{1}{|c|}{$E_{\gamma}$=1203 MeV} \\ \hline

 -0.95 &  0.122 $\pm$  0.014 &  -0.95 &  0.105 $\pm$  0.014 &  -0.95 &  0.103 $\pm$  0.014 &  -0.95 &  0.090 $\pm$  0.009 \\
 -0.85 &  0.153 $\pm$  0.020 &  -0.85 &  0.119 $\pm$  0.018 &  -0.85 &  0.110 $\pm$  0.016 &  -0.85 &  0.116 $\pm$  0.016 \\
 -0.75 &  0.175 $\pm$  0.018 &  -0.75 &  0.136 $\pm$  0.017 &  -0.75 &  0.133 $\pm$  0.017 &  -0.75 &  0.122 $\pm$  0.016 \\
 -0.65 &  0.170 $\pm$  0.017 &  -0.65 &  0.166 $\pm$  0.017 &  -0.65 &  0.157 $\pm$  0.014 &  -0.65 &  0.138 $\pm$  0.016 \\
 -0.55 &  0.199 $\pm$  0.022 &  -0.55 &  0.175 $\pm$  0.017 &  -0.55 &  0.172 $\pm$  0.018 &  -0.55 &  0.167 $\pm$  0.018 \\
 -0.45 &  0.232 $\pm$  0.029 &  -0.45 &  0.219 $\pm$  0.022 &  -0.45 &  0.184 $\pm$  0.021 &  -0.45 &  0.181 $\pm$  0.017 \\
 -0.35 &  0.250 $\pm$  0.024 &  -0.35 &  0.231 $\pm$  0.024 &  -0.35 &  0.220 $\pm$  0.023 &  -0.35 &  0.228 $\pm$  0.025 \\
 -0.25 &  0.283 $\pm$  0.026 &  -0.25 &  0.249 $\pm$  0.023 &  -0.25 &  0.254 $\pm$  0.023 &  -0.25 &  0.249 $\pm$  0.031 \\
 -0.15 &  0.272 $\pm$  0.024 &  -0.15 &  0.241 $\pm$  0.022 &  -0.15 &  0.274 $\pm$  0.025 &  -0.15 &  0.234 $\pm$  0.022 \\
 -0.05 &  0.298 $\pm$  0.025 &  -0.05 &  0.258 $\pm$  0.020 &  -0.05 &  0.275 $\pm$  0.026 &  -0.05 &  0.264 $\pm$  0.026 \\
  0.05 &  0.298 $\pm$  0.024 &   0.05 &  0.333 $\pm$  0.028 &   0.05 &  0.297 $\pm$  0.027 &   0.05 &  0.265 $\pm$  0.023 \\
  0.15 &  0.317 $\pm$  0.030 &   0.15 &  0.284 $\pm$  0.024 &   0.15 &  0.286 $\pm$  0.027 &   0.15 &  0.260 $\pm$  0.026 \\
  0.25 &  0.364 $\pm$  0.032 &   0.25 &  0.297 $\pm$  0.032 &   0.25 &  0.320 $\pm$  0.036 &   0.25 &  0.318 $\pm$  0.038 \\
  0.35 &  0.325 $\pm$  0.035 &   0.35 &  0.342 $\pm$  0.038 &   0.35 &  0.301 $\pm$  0.032 &   0.35 &  0.296 $\pm$  0.039 \\
  0.45 &  0.332 $\pm$  0.032 &   0.45 &  0.355 $\pm$  0.038 &   0.45 &  0.310 $\pm$  0.032 &   0.45 &  0.345 $\pm$  0.041 \\
  0.55 &  0.345 $\pm$  0.039 &   0.55 &  0.318 $\pm$  0.037 &   0.55 &  0.357 $\pm$  0.044 &   0.55 &  0.340 $\pm$  0.043 \\
  0.65 &  0.343 $\pm$  0.041 &   0.65 &  0.356 $\pm$  0.046 &   0.65 &  0.322 $\pm$  0.038 &   0.65 &  0.326 $\pm$  0.053 \\
  0.75 &  0.361 $\pm$  0.051 &   0.75 &  0.304 $\pm$  0.049 &   0.75 &  0.312 $\pm$  0.054 &   0.75 &  0.335 $\pm$  0.055 \\
  0.85 &  0.325 $\pm$  0.060 &   0.85 &  0.253 $\pm$  0.054 &   0.85 &  0.322 $\pm$  0.065 &   0.85 &  0.295 $\pm$  0.075 \\ \hline

\end{tabular}

\end{center}
\end{table}

\newpage

\onecolumn

\begin{table}
\begin{center}
\caption{Differential cross section $d\sigma/d\Omega$ ($\mu b/sr$) as a function of the photon laboratory energy (1200-1500 MeV) 
and the cosine of the $\eta$ center-of-mass angle.}
\label{table_sect2}

\vspace{0.3cm}

\begin{tabular}{|r|r|r|r|r|r|r|r|} \hline

\multicolumn{1}{|c|}{$\cos(\theta_{cm})$} & \multicolumn{1}{|c|}{$E_{\gamma}$=1232 MeV} &
\multicolumn{1}{|c|}{$\cos(\theta_{cm})$} & \multicolumn{1}{|c|}{$E_{\gamma}$=1261 MeV} &
\multicolumn{1}{|c|}{$\cos(\theta_{cm})$} & \multicolumn{1}{|c|}{$E_{\gamma}$=1289 MeV} &
\multicolumn{1}{|c|}{$\cos(\theta_{cm})$} & \multicolumn{1}{|c|}{$E_{\gamma}$=1317 MeV} \\ \hline

 -0.95 &  0.080 $\pm$  0.010 &  -0.95 &  0.086 $\pm$  0.012 &  -0.95 &  0.078 $\pm$  0.012 &  -0.95 &  0.076 $\pm$  0.013 \\
 -0.85 &  0.105 $\pm$  0.013 &  -0.85 &  0.103 $\pm$  0.018 &  -0.85 &  0.102 $\pm$  0.012 &  -0.85 &  0.100 $\pm$  0.016 \\
 -0.75 &  0.121 $\pm$  0.014 &  -0.75 &  0.120 $\pm$  0.016 &  -0.75 &  0.123 $\pm$  0.015 &  -0.75 &  0.106 $\pm$  0.014 \\
 -0.65 &  0.139 $\pm$  0.012 &  -0.65 &  0.143 $\pm$  0.014 &  -0.65 &  0.127 $\pm$  0.013 &  -0.65 &  0.131 $\pm$  0.014 \\
 -0.55 &  0.164 $\pm$  0.020 &  -0.55 &  0.155 $\pm$  0.015 &  -0.55 &  0.142 $\pm$  0.016 &  -0.55 &  0.129 $\pm$  0.017 \\
 -0.45 &  0.168 $\pm$  0.019 &  -0.45 &  0.181 $\pm$  0.021 &  -0.45 &  0.178 $\pm$  0.024 &  -0.45 &  0.169 $\pm$  0.023 \\
 -0.35 &  0.206 $\pm$  0.022 &  -0.35 &  0.204 $\pm$  0.025 &  -0.35 &  0.208 $\pm$  0.023 &  -0.35 &  0.186 $\pm$  0.028 \\
 -0.25 &  0.200 $\pm$  0.021 &  -0.25 &  0.212 $\pm$  0.024 &  -0.25 &  0.217 $\pm$  0.024 &  -0.25 &  0.193 $\pm$  0.021 \\
 -0.15 &  0.229 $\pm$  0.022 &  -0.15 &  0.229 $\pm$  0.023 &  -0.15 &  0.226 $\pm$  0.027 &  -0.15 &  0.213 $\pm$  0.022 \\
 -0.05 &  0.236 $\pm$  0.024 &  -0.05 &  0.216 $\pm$  0.022 &  -0.05 &  0.206 $\pm$  0.024 &  -0.05 &  0.235 $\pm$  0.026 \\
  0.05 &  0.239 $\pm$  0.023 &   0.05 &  0.258 $\pm$  0.027 &   0.05 &  0.235 $\pm$  0.024 &   0.05 &  0.231 $\pm$  0.028 \\
  0.15 &  0.264 $\pm$  0.026 &   0.15 &  0.305 $\pm$  0.035 &   0.15 &  0.272 $\pm$  0.032 &   0.15 &  0.240 $\pm$  0.036 \\
  0.25 &  0.282 $\pm$  0.033 &   0.25 &  0.294 $\pm$  0.043 &   0.25 &  0.251 $\pm$  0.040 &   0.25 &  0.269 $\pm$  0.049 \\
  0.35 &  0.306 $\pm$  0.037 &   0.35 &  0.267 $\pm$  0.038 &   0.35 &  0.270 $\pm$  0.039 &   0.35 &  0.307 $\pm$  0.038 \\
  0.45 &  0.285 $\pm$  0.043 &   0.45 &  0.312 $\pm$  0.041 &   0.45 &  0.322 $\pm$  0.047 &   0.45 &  0.261 $\pm$  0.045 \\
  0.55 &  0.330 $\pm$  0.042 &   0.55 &  0.333 $\pm$  0.044 &   0.55 &  0.278 $\pm$  0.039 &   0.55 &  0.302 $\pm$  0.052 \\
  0.65 &  0.344 $\pm$  0.050 &   0.65 &  0.336 $\pm$  0.062 &   0.65 &  0.335 $\pm$  0.062 &   0.65 &  0.302 $\pm$  0.057 \\
  0.75 &  0.360 $\pm$  0.089 &   0.75 &  0.303 $\pm$  0.060 &   0.75 &  0.308 $\pm$  0.069 &   0.75 &  0.299 $\pm$  0.068 \\
  0.85 &  0.198 $\pm$  0.051 &   0.85 &  0.245 $\pm$  0.071 &        &                    &        &                    \\ \hline

\end{tabular}

\begin{tabular}{|r|r|r|r|r|r|r|r|} \hline

\multicolumn{1}{|c|}{$\cos(\theta_{cm})$} & \multicolumn{1}{|c|}{$E_{\gamma}$=1344 MeV} &
\multicolumn{1}{|c|}{$\cos(\theta_{cm})$} & \multicolumn{1}{|c|}{$E_{\gamma}$=1371 MeV} &
\multicolumn{1}{|c|}{$\cos(\theta_{cm})$} & \multicolumn{1}{|c|}{$E_{\gamma}$=1399 MeV} &
\multicolumn{1}{|c|}{$\cos(\theta_{cm})$} & \multicolumn{1}{|c|}{$E_{\gamma}$=1425 MeV} \\ \hline

 -0.95 &  0.076 $\pm$  0.013 &  -0.95 &  0.068 $\pm$  0.012 &  -0.95 &  0.068 $\pm$  0.012 &  -0.95 &  0.065 $\pm$  0.013 \\
 -0.85 &  0.090 $\pm$  0.012 &  -0.85 &  0.093 $\pm$  0.015 &  -0.85 &  0.076 $\pm$  0.015 &  -0.85 &  0.080 $\pm$  0.014 \\
 -0.75 &  0.108 $\pm$  0.017 &  -0.75 &  0.100 $\pm$  0.015 &  -0.75 &  0.103 $\pm$  0.014 &  -0.75 &  0.087 $\pm$  0.015 \\
 -0.65 &  0.112 $\pm$  0.014 &  -0.65 &  0.112 $\pm$  0.012 &  -0.65 &  0.099 $\pm$  0.012 &  -0.65 &  0.094 $\pm$  0.014 \\
 -0.55 &  0.139 $\pm$  0.017 &  -0.55 &  0.120 $\pm$  0.011 &  -0.55 &  0.113 $\pm$  0.013 &  -0.55 &  0.104 $\pm$  0.015 \\
 -0.45 &  0.162 $\pm$  0.028 &  -0.45 &  0.173 $\pm$  0.026 &  -0.45 &  0.155 $\pm$  0.023 &  -0.45 &  0.132 $\pm$  0.021 \\
 -0.35 &  0.211 $\pm$  0.030 &  -0.35 &  0.175 $\pm$  0.027 &  -0.35 &  0.166 $\pm$  0.028 &  -0.35 &  0.164 $\pm$  0.030 \\
 -0.25 &  0.191 $\pm$  0.025 &  -0.25 &  0.177 $\pm$  0.023 &  -0.25 &  0.181 $\pm$  0.027 &  -0.25 &  0.158 $\pm$  0.022 \\
 -0.15 &  0.197 $\pm$  0.025 &  -0.15 &  0.188 $\pm$  0.023 &  -0.15 &  0.190 $\pm$  0.021 &  -0.15 &  0.160 $\pm$  0.024 \\
 -0.05 &  0.222 $\pm$  0.022 &  -0.05 &  0.188 $\pm$  0.025 &  -0.05 &  0.200 $\pm$  0.027 &  -0.05 &  0.185 $\pm$  0.024 \\
  0.05 &  0.221 $\pm$  0.029 &   0.05 &  0.239 $\pm$  0.031 &   0.05 &  0.185 $\pm$  0.024 &   0.05 &  0.206 $\pm$  0.032 \\
  0.15 &  0.239 $\pm$  0.038 &   0.15 &  0.207 $\pm$  0.031 &   0.15 &  0.220 $\pm$  0.034 &   0.15 &  0.232 $\pm$  0.040 \\
  0.25 &  0.270 $\pm$  0.048 &   0.25 &  0.258 $\pm$  0.041 &   0.25 &  0.239 $\pm$  0.044 &   0.25 &  0.193 $\pm$  0.040 \\
  0.35 &  0.265 $\pm$  0.043 &   0.35 &  0.263 $\pm$  0.041 &   0.35 &  0.217 $\pm$  0.042 &   0.35 &  0.247 $\pm$  0.047 \\
  0.45 &  0.250 $\pm$  0.042 &   0.45 &  0.252 $\pm$  0.039 &   0.45 &  0.293 $\pm$  0.064 &   0.45 &  0.222 $\pm$  0.039 \\
  0.55 &  0.264 $\pm$  0.051 &   0.55 &  0.233 $\pm$  0.047 &   0.55 &  0.253 $\pm$  0.049 &   0.55 &  0.266 $\pm$  0.053 \\
  0.65 &  0.286 $\pm$  0.059 &   0.65 &  0.268 $\pm$  0.061 &   0.65 &  0.281 $\pm$  0.064 &   0.65 &  0.245 $\pm$  0.051 \\
  0.75 &  0.298 $\pm$  0.064 &   0.75 &  0.269 $\pm$  0.060 &   0.75 &  0.263 $\pm$  0.080 &   0.75 &  0.267 $\pm$  0.070 \\ \hline

\end{tabular}

\begin{tabular}{|r|r|r|r|} \hline
  
\multicolumn{1}{|c|}{$\cos(\theta_{cm})$} & \multicolumn{1}{|c|}{$E_{\gamma}$=1450 MeV} &
\multicolumn{1}{|c|}{$\cos(\theta_{cm})$} & \multicolumn{1}{|c|}{$E_{\gamma}$=1477 MeV} \\ \hline

 -0.95 &  0.050 $\pm$  0.013 &  -0.95 &  0.034 $\pm$  0.016 \\ 
 -0.85 &  0.068 $\pm$  0.016 &  -0.85 &  0.053 $\pm$  0.016 \\ 
 -0.75 &  0.077 $\pm$  0.013 &  -0.75 &  0.054 $\pm$  0.015 \\ 
 -0.65 &  0.092 $\pm$  0.010 &  -0.65 &  0.087 $\pm$  0.018 \\ 
 -0.55 &  0.081 $\pm$  0.016 &  -0.55 &  0.058 $\pm$  0.012 \\ 
 -0.45 &  0.099 $\pm$  0.017 &  -0.45 &  0.093 $\pm$  0.027 \\ 
 -0.35 &  0.136 $\pm$  0.026 &  -0.35 &  0.130 $\pm$  0.031 \\
 -0.25 &  0.134 $\pm$  0.032 &  -0.25 &  0.148 $\pm$  0.032 \\
 -0.15 &  0.139 $\pm$  0.021 &  -0.15 &  0.110 $\pm$  0.022 \\
 -0.05 &  0.134 $\pm$  0.025 &  -0.05 &  0.156 $\pm$  0.036 \\
  0.05 &  0.179 $\pm$  0.031 &   0.05 &  0.201 $\pm$  0.038 \\
  0.15 &  0.174 $\pm$  0.038 &   0.15 &  0.164 $\pm$  0.051 \\
  0.25 &  0.170 $\pm$  0.041 &   0.25 &  0.142 $\pm$  0.041 \\
  0.35 &  0.225 $\pm$  0.052 &   0.35 &  0.144 $\pm$  0.038 \\
  0.45 &  0.227 $\pm$  0.052 &   0.45 &  0.251 $\pm$  0.064 \\
  0.55 &  0.208 $\pm$  0.048 &   0.55 &  0.164 $\pm$  0.054 \\
  0.65 &  0.214 $\pm$  0.062 &   0.65 &  0.177 $\pm$  0.064 \\ 
  0.75 &  0.213 $\pm$  0.093 &        &                     \\ \hline

\end{tabular}

\end{center}
\end{table}

\end{document}